\documentclass[pre,reprint,superscriptaddress]{revtex4-1}
\usepackage[usenames,dvipsnames,svgnames,table]{xcolor}

\usepackage{float}
\usepackage{amsmath}
\usepackage{amssymb}
\usepackage{graphicx}
\usepackage{hyperref}
\newcommand{\changes}[1]{{\color{black} #1}} 
\newcommand{\changess}[1]{{\color{black} #1}} 
\newcommand{\changesss}[1]{{\color{black} #1}} 

\makeatletter

\usepackage[caption=false]{subfig}

\makeatother

\begin{document}

\title{The Marangoni effect on small-amplitude capillary waves in
viscous fluids}

\author{Li \surname{Shen}}
\thanks{Corresponding Author}
\email{l.shen14@imperial.ac.uk}

\author{Fabian Denner}
\affiliation{Department of Mechanical Engineering, Imperial College London, London,
SW7 2AZ, United Kingdom}
\author{Neal Morgan}
\affiliation{Shell Global Solutions Ltd, Brabazon House, Threapwood Road, Manchester, M22 0RR, United Kingdom}
\author{Berend van Wachem}
\affiliation{Department of Mechanical Engineering, Imperial College London, London,
SW7 2AZ, United Kingdom}
\author{Daniele Dini}
\affiliation{Department of Mechanical Engineering, Imperial College London, London,
SW7 2AZ, United Kingdom}

\date{\today}
\begin{abstract}
We derive a general integrodifferential equation for the transient behaviour
of small-amplitude capillary waves on the planar surface of a viscous
fluid in the presence of the Marangoni effect. \changesss{The equation is solved for an insoluble surfactant solution in concentration below the critical micelle concentration (cmc) undergoing convective-diffusive surface transport. The special case of a diffusion-driven surfactant is considered near the the critical damping wavelength.} The Marangoni effect is shown to contribute to the overall damping mechanism and a first-order term correction to the critical wavelength with respect to the surfactant concentration difference and the Schmidt number is proposed.
\end{abstract}
\maketitle

\section{Introduction}

On the free surface between any two immiscible phases, external perturbations such as \changes{thermally-induced} motion or vibrations  can \changes{cause}
statistical fluctuations \cite{Aarts2004}, or interfacial waves, leading to rough surfaces \changes{\cite{Aarts2004,Mandelstam1913}}. Long waves with wavelength $\lambda \gg \sqrt{\sigma/(\rho g)}\equiv l_*$ (where $\rho$ is fluid density, $g$ is gravitational acceleration and $\sigma$ is surface tension \changes{coefficient}) are gravity-driven and known as gravity waves. Short waves with wavelength $\lambda \ll l_*$ are \textit{capillary waves} where the  restoring effects of surface tension dominate gravity. Furthermore, in regions of \changes{large} viscosity, the energy dissipation is also affected by the distribution of fluid vorticity \cite{Prosperetti1976} leading to an increasing viscous attenuation of the capillary wave \cite{Lamb1932} for decreasing wavelength. The evolution of such waves is critical to the study of film stability, e.g. thin film rupture \cite{Scheludko1967}, droplet coalescence \cite{Blanchette2006}, polymer films \cite{Sferrazza1997} and foam dynamics \cite{Saye2013} as well as capillary-driven phenomena, such as the breakup of liquid sheets, jets \cite{Papageorgiou1993,Eggers2008}, bridges \cite{Hoepffner2013}, ligaments \cite{Eggers2008}  and curtains \cite{Lhuissier2016}\changes{.}

In addition, interfaces \changes{found in nature and engineering applications are often} contaminated with surface-active substances. These surfactants can significantly alter interfacial flow structures and energy dissipation through the Marangoni effect as a result of the generated gradients in surface tension \cite{Batchelor2003,Karakashev2014}. Hence, an understanding of how the Marangoni effect changes the damping of the interface through the attenuation of capillary waves could extend the classification of surface roughness \cite{Mandelstam1913} to  a greater number of industrial and biological films and membranes. 

The classical treatment of the capillary wave considers the interfacial
displacement as an infinite superposition of modes, then assuming
each component contributes according to an equipartition theorem \cite{Aarts2004,Sides1999} or approximating the solution as the least-damped mode \cite{Lamb1932,Chandrasekhar1981}.
In either cases, whilst the long-time behaviour is sufficiently captured
as the waves are damped out and converge towards a single mode, the transient
behaviour remains difficult to extract \changes{because of} its dependence on multiple of such
modes. To circumvent this difficulty, Prosperetti \cite{Prosperetti1976}
recast the equation of motion for viscous fluids as an integrodifferential
equation solved exactly as an initial value problem. The long-time connection of the solution with the normal mode approach \changes{was also demonstrated}
through the complex characteristic equation for the discrete wave
spectrum \cite{Lamb1932} along with the equivalence of irrotational
flows with the short-time approximation of the solution. Motivated by this approach,
we consider in this paper the capillary waves in viscous fluids with a
\changes{significant} Marangoni effect \changes{in} the presence of a surfactant solution. 

As a prelude to our discussion, we consider the implicit dispersion
relation of the capillary wave 
\begin{equation}
\omega_{0}^{2}+(\mathrm{i}\omega+2\nu k^2)^{2}-4\nu^2 k^4\left(1+\frac{\mathrm{i}\omega}{\nu k^2}\right)^{1/2}=0,\label{eq:dispersion}
\end{equation}
\changes{derived by} linear response theory or linearised hydrodynamics \cite{Delgado2008a,Jackle1999} under the assumption of weak viscous damping,
where $\nu$ is the kinematic viscosity, $k$ is wavenumber, $\omega$ is angular frequency of the capillary wave  and  
\begin{equation}
\omega_0=\left(\frac{\sigma k^{3}}{\rho}+gk\right)^{1/2}
\end{equation}
is the \changes{angular frequency} on an interface of an inviscid ideal fluid. For large wavenumber $k\gg 1$, $\omega_0\simeq (\sigma k^3/\rho)^{1/2}$, \changes{at which} the effects of gravity is negligible. By inspection, Eq.\,(\ref{eq:dispersion}) suggests a critical wavenumber $k_{c}$ 
where there is a transition from the propagative and underdamped mode to an overdamped mode \cite{Delgado2008a}
in which the effect of viscosity dominates the oscillatory motion of the interface. More recently, Denner \cite{Denner2016b} proposed a rational parametrisation for the dispersion in the underdamped regime, including critical damping, that leads to a self-similar solution of the frequency dispersion of capillary waves in viscous fluids. 

The region near the critical wavelength $\lambda_\mathrm{c}=2\pi/k_\mathrm{c}$ is particularly interesting due to its proximity to the threshold \changes{whereby} complex rupture behaviours are exhibited by biological and industrial interfacial systems \cite{Karakashev2014,Scheludko1967,Yaminsky2010,Cantat2013}. Therefore, the evolution of $\lambda_\mathrm{c}$ \changes{as a result of} the Marangoni effect could potentially provide clues to the onset of these \changes{phenomena.}

For constant surface tension, the wavenumber $k_\mathrm{c}$ can be related on dimensional grounds to the quantity $\delta_{\nu}=(\nu/\omega)^{1/2}$,
which is the characteristic thickness of the penetration of vorticity
that is generated at the interface boundary due to the inability of
the initially irrotational flow to sustain the non-zero tangential
boundary stress condition \cite{Batchelor2000}. \changess{More precisely, it has been shown \cite{Denner2016b,Denner2016c},  that $k_\text{c}\sim 1/l_\text{vc}$ where 
\begin{equation}
	l_\text{vc}=\frac{\mu^2}{\rho\sigma}
\end{equation}
is a viscocapillary length scale under the regime $\mathrm{Oh}=1$, where
\begin{equation}
	\mathrm{Oh}=\frac{\mu}{\sqrt{\sigma\rho\, l}}=\frac{t_\sigma}{t_\mu}
\end{equation}
is the  Ohnesorge number. The quantity $t_\sigma=\sqrt{\rho\, l^3/\sigma}$ is the capillary timescale proportional to the period of an undamped capillary wave and  $t_\mu=\rho\, l^2/\mu$ is the viscous timescale associated with the length of time taken for momentum to diffuse through a characteristic lengthscale $l$.}

In the presence of a variable surface tension, $\lambda_\mathrm{c}$ must increase since the motion driven by surface tension gradients is always dissipative in nature due to the induced viscous shear stresses. Consider the stress boundary condition 
\begin{equation}
-[\mathbf{T}\cdot \mathbf{n}]+2H\sigma \mathbf{n} = M\changes{,}	
\end{equation}
where $\mathbf{T}$ is the viscous stress tensor, $\mathbf{n}$ the normal vector, $2H=-\boldsymbol{\nabla}_s\cdot\mathbf{n}$ the mean curvature; where 
\begin{equation}
	\boldsymbol{\nabla}_s=\mathbf{(I-nn)\cdot\boldsymbol{\nabla}}
\end{equation}
is the surface gradient operator; $M$ is the Marangoni term and $[\cdot]$ denotes the jump across the interface of the quantity within. For systems with surface-active substances, Levich and Krylov \cite{Levich1969} suggested $M=-\mu_s \nabla^2 \mathbf{u}_s$ where $\mathbf{u}_s=\mathbf{(I-nn)\cdot\mathbf{u}}$ is the surface velocity and $\mu_s$ is a phenomenological surface viscosity coefficient. This term explicitly accounts for the energy dissipation at the interface due to the irreversible processes involved and hence a natural consequence is that the Marangoni effect directly contributes to the overall damping of the system. However, the phenomenological and model-specific nature of $\mu_s$ does not easily yield to general analysis and so in this paper we shall adopt the classical approach commonly used in literature \cite{Batchelor2003} of letting $M=-\boldsymbol{\nabla}_s\sigma$, coupled with an equation of state linking surface tension $\sigma$ and surfactant concentrations (or another field such as temperature).  

Specialising to the case of a predominantly-diffusive surfactant solution, the critical wavenumber $k_\mathrm{c}$  is now dimensionally related to the ratio $\delta_{\nu}/\delta_{D}$, 
where $\delta_{D}=(D/\omega)^{1/2}$ is the thickness of the mass transfer boundary
layer with \changes{surfactant} diffusivity $D$, more commonly known as its square the
Schmidt number, 
\begin{equation}
	\mathrm{Sc}=\left(\frac{\delta_{\nu}}{\delta_{D}}\right)^{2}=\frac{\nu}{D}. 
\end{equation}

\section{Equation of motion}

The dynamics of the fluid of viscosity $\mu$, and density $\rho$
in regions of Reynolds number $\mathrm{Re}\ll1$ are described by
the \changes{time-dependent} Stokes' equation 

\begin{alignat}{1}
\frac{\partial\mathbf{u}}{\partial t} & =-\frac{1}{\rho}\boldsymbol{\nabla} p+\nu\nabla^2\mathbf{u}-g\mathbf{j}\\
\boldsymbol{\nabla}\cdot\mathbf{u} & =0
\end{alignat}
where $\mathbf{u}$ is the 2-dimensional fluid velocity field $(u,v)$, $p$ is the
pressure and $g\mathbf{j}$ is the gravity term with $\mathbf{j}$
denoting the upward unit vector. The free surface is given by the standing
wave 
\begin{alignat}{1}
F(x,y,t) & =y-\mathrm{f}(x,t)\\
\mathrm{f}(x,t) & =a(t)\cos kx
\end{alignat}
where the non-linear, time-dependent wave amplitude $a(t)$ satisfies the small-amplitude conditions that 
\begin{alignat}{2}
	a       & \ll \lambda & &= \frac{2\pi}{k},\\
	\frac{\mathrm{d}a}{\mathrm{d}t} & \ll v_\mathrm{p} & &=\frac{\omega}{k},
\end{alignat}
for wavelength $\lambda$, wavenumber $k$, angular frequency $\omega$ and phase velocity $v_\mathrm{p}$. The linearised tangential and normal stress components, \changess{$\mathrm{T}_\parallel$ and $\mathrm{T}_\perp$, respectively,} as well
as the kinematic conditions are given on the surface $y=0$ by 
\begin{alignat}{2}
\mathrm{T}_{\parallel}\equiv & \,\frac{1}{2}\mu\left(\frac{\partial v}{\partial x}+\frac{\partial u}{\partial y}\right) &  & =\mathbf{t}\cdot\boldsymbol{\nabla}\sigma,\label{eq:tangential}\\
\mathrm{T}_{\perp}\equiv & -p+2\mu\frac{\partial v}{\partial y} &  & =\sigma\boldsymbol{\nabla}_s\cdot\mathbf{n},\label{eq:normal}\\
 & \frac{\partial F}{\partial t}+v\frac{\partial F}{\partial y} &  & =0,\label{eq:kinematic}
\end{alignat}
where $\mathbf{n}\simeq(ak\sin kx,1)$ and $\mathbf{t}\simeq(1,-ak\sin kx)$
are the leading order normal and tangent vectors. The surface tension coefficient $\sigma$
is modelled by a linear equation of state 
\begin{equation}
\sigma=\sigma_{0}-\alpha\Gamma(x,y,t),
\end{equation}
where $\alpha=\left|\mathrm{d}\sigma/\mathrm{d}\Gamma(x,y,t)\right|$ is
the magnitude of surface tension gradient and $\Gamma(x,y,t)$
is a function which depends on the mode
of the Marangoni effect (e.g. thermal, soluto or electric etc, \cite{Karakashev2014}.) \changes{In this paper, we shall consider the soluto-Marangoni effect of an interface in the presence of a surfactant solution. Henceforth, $\Gamma=\Gamma(x,t)$ is the surface concentration of an insoluble surfactant solution and $D=D_s$ denotes the surface diffusivity coefficient.}

Following \cite{Prosperetti1976}, we decompose the velocity
and pressure into harmonic and viscous corrections 
\begin{equation}
	(\mathbf{u},\,p)=(\mathbf{u}'+\mathbf{u}'',\,p'+p''),
\end{equation}
where the harmonic component $(\mathbf{u}',p')$ assumes the velocity field 
\begin{equation}
	\mathbf{u}'=(u',v')=\boldsymbol{\nabla}\phi
\end{equation}
for a harmonic function $\phi$ satisfying the hydrostatic potential
problem 
\begin{alignat}{1}
\frac{\partial\mathbf{u}'}{\partial t} & =-\frac{1}{\rho}\boldsymbol{\nabla} p'-g\mathbf{j}\\
 0 & =\left(\frac{\partial F}{\partial t}+v'\frac{\partial F}{\partial y}\right)_{y=0},
\end{alignat}
with known solution \cite{Lamb1932,Prosperetti1976}
\begin{alignat}{1}
\phi & =\frac{1}{k}\frac{\mathrm{d}a}{\mathrm{d}t}\mathrm{e}^{ky}\cos kx,\\
p' & =-\rho\left(gy+\frac{1}{k}\frac{\mathrm{d}^{2}a}{\mathrm{d}t^{2}}\mathrm{e}^{ky}\cos kx\right).
\end{alignat}
On the other hand, the incompressible viscous correction component $(\mathbf{u}'',p'')$ satisfies the Stokes' problem
\begin{alignat}{1}
\frac{\partial\mathbf{u}''}{\partial t} & =-\frac{1}{\rho}\boldsymbol{\nabla} p''+\nu\nabla^2\mathbf{u}''\label{eq:viscous}\\
0 & =\left.v''\frac{\partial F}{\partial y}\right|_{y=0}
\end{alignat}
and the incompressibility condition $\nabla\cdot\mathbf{u}''=0$. 

Taking the curl of Eq.$\,$(\ref{eq:viscous}) yields the vorticity equation 
\begin{equation}
\frac{\partial \omega_z}{\partial t}=\nu\nabla^2\omega_{z},\label{eq:vorticity}
\end{equation}
where $\omega_{z}$ is given on the free surface $\mathrm{f}(x,t)=a\cos kx$ by
\begin{alignat}{1}
\omega_{z}(x,f,t) & =2\left(-\frac{\mathrm{T}_{\parallel}}{\mu}+\frac{\partial v}{\partial x}\right)\\
 & =2\left(-\frac{\mathrm{T}_{\parallel}}{\mu}+\frac{\partial^{2}f}{\partial x\partial t}\right)
\end{alignat}
using the kinematic condition 
\begin{equation}
0=\frac{\partial F}{\partial t}+(\mathbf{u}\cdot\boldsymbol{\nabla})F\simeq-\frac{\partial f}{\partial t}+v
\end{equation}
and the insoluble surfactant condition 
\begin{equation}
	\mathbf{t}\cdot\boldsymbol{\nabla}\sigma=\frac{\partial\sigma}{\partial x},
\end{equation}
which neglects any surface adsorption processes.

Satisfying Eq.$\,$(\ref{eq:viscous}) with the stream-function $\psi$ defined by
\begin{equation}
\left(u'',\,v''\right)=\left(\frac{\partial\psi}{\partial y},\,-\frac{\partial\psi}{\partial x}\right),
\end{equation}
we can write 
\begin{equation}
	\omega_{z}=\nabla^{2}\psi
\end{equation}
such that Eq.$\,$(16) takes
on the bi-harmonic form 
\begin{equation}
\frac{\partial(\nabla^{2}\psi)}{\partial t}=\nabla^{4}\psi.\label{eq:bihar}
\end{equation}
Furthermore, writing 
\begin{equation}
\left(\omega_{z},\,\psi\right)=\left(\Omega(y,t),\,\Psi(y,t)\right)\sin kx,
\end{equation}
and assuming that $\Psi$ is bounded as $y\rightarrow-\infty$ gives
the Green's function solution of the form  
\begin{alignat}{1}
2k\Psi & =\mathrm{e}^{ky}\left(\int_{-\infty}^{0}\Omega\mathrm{e}^{k\xi}\mathrm{d}\xi+\int_{0}^{y}\Omega\mathrm{e}^{-k\xi}\mathrm{d}\xi\right)\label{eq:psi1}\\
 & \qquad\qquad\qquad\quad-\mathrm{e}^{-ky}\int_{-\infty}^{y}\Omega\mathrm{e}^{k\xi}\mathrm{d}\xi\changes{,}\nonumber 
\end{alignat}
where $\Omega$ satisfies the second-order equation 
\begin{equation}
\frac{\partial \Omega}{\partial t}=\nu\left(\frac{\partial \Omega}{\partial y^2}-k^{2}\Omega\right),\label{eq:Omega}
\end{equation}
with the initial condition $\Omega(y,0)=0.$ 

\changes{Writing the surfactant concentration in the wave form
\begin{equation}
 	\Gamma(x,t)-\Gamma_0=\tilde{\Gamma}(t)\cos kx
 \end{equation}} 
then yields the boundary condition
\begin{equation}
\Omega(0,t)=-2k\left(\frac{\alpha\tilde\Gamma(t)}{\mu}+\frac{\mathrm{d}a}{\mathrm{d}t}\right).
\end{equation}
\changes{The} Laplace transform of Eq.\,(\ref{eq:Omega}) then gives 
\begin{equation}
	\hat{\Omega}(y,s)=-2k\mathcal{L}\left[\frac{\alpha \tilde{\Gamma}}{\mu}+\frac{\mathrm{d}a}{\mathrm{d}t}\right]\mathrm{e}^{{y(k^{2}+s/\nu)^{1/2}}},
\end{equation}
where 
\begin{equation}
\hat{f}(s)\equiv\mathcal{L}\left[f(t)\right]=\int^\infty_0f(t)\mathrm{e}^{-ts}\mathrm{d}t
\end{equation}
is the Laplace transform of $f(t)$. 

Integrating Eq.$\,$(\ref{eq:psi1}), eliminating $\partial \Omega/\partial t$ using Eq.$\,$(\ref{eq:Omega}) and integration by parts yield 
\begin{equation}
\frac{\partial\Psi}{\partial t}=\nu[\Omega(y,t)-\Omega(0,t)\mathrm{e}^{ky}]
\end{equation}
and the viscous pressure is given by  
\begin{equation}
	p''=\mu\Omega(0,t)\mathrm{e}^{ky}\cos kx.
\end{equation}

Substituting Eq.$\,$(8) into the normal stress condition gives the
integro-differential equation of motion in the non-dimensional form

\begin{widetext}
\begin{alignat}{1}
 & \frac{\mathrm{d}^{2}a}{\mathrm{d}\tau^{2}}+\epsilon\left(4\frac{\mathrm{d}a}{\mathrm{d}\tau}+2\beta\tilde{\Gamma}(\tau)\right)+a-4\epsilon^{2}\int_{0}^{\tau}\left(\beta\tilde{\Gamma}(\tau')+\frac{\mathrm{d}a(\tau')}{\mathrm{d}\tau'}\right)\left(\frac{1}{\sqrt{\epsilon\pi(\tau-\tau')}}\mathrm{e}^{-\epsilon(\tau-\tau')}-\mathrm{erfc}\sqrt{\epsilon(\tau-\tau')}\right)\mathrm{d}\tau'=0,\label{eq:EoM}
\end{alignat}

\end{widetext}where 
\begin{equation}
	\mathrm{erfc}(z)=\frac{2}{\sqrt{\pi}}\int_{z}^{\infty}\mathrm{e}^{-y^{2}}\mathrm{d}y
\end{equation}
is the complementary error function, $\tau=\omega t$ is the non-dimensional
time, $\tau'$ is the integration variable (a non-dimensional time discussed in sec.\,III), $\epsilon=\nu k^{2}/\omega$ is the non-dimensional kinematic viscosity
and the non-dimensional surfactant diffusivity and surfactant strength
parameters are given by 
\begin{equation}
\left(\zeta,\beta\right)=\frac{k}{\omega}\left(Dk,\frac{\alpha\Gamma_{0}}{\mu}\right).
\end{equation}

In the absence of the Marangoni effect, i.e. for $\beta=0$, Eq.$\,$(\ref{eq:EoM})
reduces to 
\begin{equation}
	\ddot{a}+4\epsilon\dot{a}+a-4\epsilon^{2}\dot{a}*\mathcal{F}(\tau)=0\label{eq:pEoM},
\end{equation}
as derived in \cite{Prosperetti1976},  where $\dot{}=\mathrm{d}/\mathrm{d}\tau$,
$*$ is the convolution operator and $\mathcal{F}(\tau)$ is the auxiliary
function 
\begin{equation}
\mathcal{F}(\tau)=\frac{1}{\sqrt{\pi\epsilon\tau}}\mathrm{e}^{-\epsilon\tau}-\mathrm{erfc}\sqrt{\epsilon\tau}.
\end{equation}

\section{Non-linear viscous dissipation via fractional integral}

The viscous equation of motion \changes{in Eq.}\,(\ref{eq:pEoM}) can be rewritten in the form 
\begin{equation}
\ddot{a}+4\epsilon \dot{a}-4\epsilon^{3/2} \mathrm{I}^{1/2}[\mathrm{G}(\tau')\mathrm{e}^{-\epsilon(\tau-\tau')}] +(1+4\epsilon^2)a= \mathrm{J}(\tau),\label{eq:EoMviscousNL}
\end{equation}
where $\mathrm{J}(\tau)$ is the forcing function given by 
\begin{equation}
\mathrm{J}(\tau)=4\epsilon^{2}a_0\mathrm{erfc}\sqrt{\epsilon\tau}
\end{equation}
and $\mathrm{G}(\tau')$ is a local viscous dissipation density function
defined by 
\begin{equation}
\mathrm{G}(\tau')=\dot{a}(\tau')+\epsilon a(\tau').\label{eq:GJ}
\end{equation}
Recall the definition for the left-handed Riemann-Liouville fractional integral \changess{of order $\varrho$}
\cite{Podlubny2001,Atanackovic2014} for $t>0,\varrho\in\mathbb{C}$ via
\begin{alignat}{1}
\mathrm{I}^{\varrho}f(\tau) & \equiv\frac{1}{\mathrm{g}(\varrho)}\int_{0}^{\tau}(\tau-\tau')^{\varrho-1}\mathrm{f}(\tau')\,\mathrm{d}\tau'.\\
 & \equiv\int_{0}^{\tau}\mathrm{f}(\tau')\,\mathrm{d}\tau_{\nu}(\tau';\tau,\varrho)\label{eq:V-Time}
\end{alignat} 
where \changes{$\mathrm{f}(\tau)$ is a function in time,} 
\begin{equation}
\mathrm{g}(\varrho)=\int^\infty_0 t^{\varrho-1}\mathrm{e}^{-t}\,\mathrm{d}t	
\end{equation}
is the gamma function and  
\begin{equation}
	\tau_{\nu}(\tau';\tau,\varrho)=\frac{\tau^{\varrho}-(\tau-\tau')^{\varrho}}{\mathrm{g}(\varrho+1)}\label{eq:inhomtime}
\end{equation}
  \changess{is a non-linear time  which satisfies the scaling property such that for the local linear time variables $\tau'=\omega t'$ and $\tau=\omega t$, we have 
\begin{equation}
\tau_\nu(\tau';\tau,\varrho)\equiv\tau_\nu(\omega t';\omega t;\varrho)=\omega^{\varrho}\tau_\nu(t';t,\varrho).
\end{equation}

}

We can now interpret the effect of viscosity on the system as introducing
a forcing term and both a linear and non-linear viscous dissipation term.
Recasting the irrotationally inviscid surface wave system into the
well-known Hamiltonian form \cite{Miles1977,Zhang1997}, the governing
equation in the fluid region $\Upsilon$ is written 
\begin{equation}
\left(\dot{h}(\mathbf{x},\tau),\dot{\Phi}(\mathbf{x},\tau)\right)=\left(\frac{\delta \mathcal{H}}{\delta\Phi},-\frac{\delta \mathcal{H}}{\delta h}+Q(h,\Phi)\right)\changes{,}	
\end{equation}
where $\mathcal{H}$ is the Hamiltonian defined by 
\begin{alignat}{1}
\mathcal{H}[h,\Phi] & = \frac{1}{2}\int_{\Upsilon}\mathrm{d}\mathbf{x}\left[\left|\boldsymbol{\nabla}\Phi\right|^{2}+gh^{2}+\sigma(\sqrt{1+|\boldsymbol{\nabla} h|^{2}}-1)\right],
\end{alignat}
and $h=h(\mathbf{x},t),\Phi=\phi(\mathbf{x},h,t)$ are the canonical coordinate, momentum variables
denoting fluid height and interfacial fluid velocity potential, respectively, for $\mathbf{x}=(x,y)$. 
The term $4\epsilon\dot{a}$ can then be viewed as the linear
part the dissipation function $Q(h,\Phi)$ of the system as we equate the rate
of loss of energy \cite{Landau1966,Zhang1997}
\begin{equation}
\frac{\mathrm{d}\mathcal{H}}{\mathrm{d}t}-\frac{\partial \mathcal{H}}{\partial t} =\int_\Upsilon \left(Q(h,\Phi)\frac{\partial h}{\partial t}\right) \mathrm{d}\mathbf{x}
\end{equation}
 with the rate of dissipation in incompressible fluids
under irrotational flow given \cite{Batchelor2000} by 
\begin{equation}
2\nu\int_\Upsilon \mathrm{d}\mathbf{x} \int^h_{-\infty}\left(\frac{\partial^2\phi}{\partial x_i \partial x_j}\right)^{2}	=\nu\int_\Upsilon \mathrm{d}\mathbf{x} \int^h_{-\infty}\nabla^2 (q^{2}),		
\end{equation}
where $q^{2}=(\partial \phi/\partial x_i)^{2}$ for velocity potential $\phi$.

The fractional integral term in Eq.\,(\ref{eq:EoMviscousNL}) suggests that the non-linear part of the dissipation function $Q(h,\Phi)$
is the exponential decay of $\mathrm{G}$ with
decay rate $\epsilon$ (which is half the rate obtained \cite{Lamb1932}
for the inviscid flow with velocity potential $\phi=k^{-1}\dot{a}\mathrm{e}^{ky}\cos kx$). \changess{Let 
\[
\mathrm{D}(\tau)=\int^\tau_0\mathrm{G}(\tau')\mathrm{e}^{-\epsilon(\tau-\tau')}\mathrm{d}\tau_\nu(\tau';\tau,\tfrac{1}{2}),
\]
then $\mathrm{D}(\tau)$ can be interpreted as the actual non-linear
viscous dissipation for which the local values of the dissipation density $\mathrm{G}(\tau')\mathrm{e}^{-\epsilon(\tau-\tau')}$ and the local linear time $\tau'$ are encoded with respect to the global inhomogeneous timescale given by the function $\tau_{\nu}(\tau';\tau,\frac{1}{2})$. From Fig.\,1, $\tau_\nu$ is approximately linear for small time $\tau$ and the exponentially-decaying fractional integral term can be shown \cite{Prosperetti1976} to be negligible. As $\tau$ increases, the timescale $\tau_\nu$ visibly slows down and deviates from the linear time. This slow-down for large $\tau$ could explain why the integral term in Eq.(\ref{eq:EoM}) cannot be neglected for $\tau'\gg \epsilon^{-1}$; since the exponential decay of the local viscous dissipation is progressively retarded for increasing $\tau$.
}

Incorporating the Marangoni effect, the equation of motion Eq.\,(\ref{eq:EoM}) takes the same form as Eq\,(\ref{eq:EoMviscousNL}) where the forcing function $\mathrm{J}$ and the local viscous dissipation density function $\mathrm{G}$ are transformed via the map $\Pi$ defined by 
\begin{equation}
	\Pi\begin{pmatrix}\mathrm{J}(\tau)\\
\mathrm{G}(\tau)
\end{pmatrix}=\begin{pmatrix}\mathrm{J}(\tau)\\
\mathrm{G}(\tau)
\end{pmatrix}+\beta\begin{pmatrix}-2\epsilon[\tilde{\Gamma}(\tau)+2\epsilon\gamma(\tau)]\\
\tilde{\Gamma}(\tau)+\epsilon\gamma(\tau)
\end{pmatrix}
\end{equation}
where 
\begin{equation}
\gamma(\tau)=\int_{0}^{\tau}\tilde{\Gamma}(\tau')\,\mathrm{d}\tau'.	
\end{equation}

\changesss{\section{Surfactant modelling: Marangoni Convection}

To model the surfactant transport, we consider a mass balance \cite{Edwards1991}
at the interface for an insoluble surfactant of concentration $\Gamma(x,t)$.
The convective-diffusive equation takes the leading-order form
\begin{equation}
\frac{\partial\Gamma(x,t)}{\partial t}+\frac{\partial}{\partial x}\left[\Gamma(x,t=0)u\right]=D\frac{\partial^{2}\Gamma}{\partial x^{2}}.\label{eq:SURF-TRANS}
\end{equation}
Assuming the waveform 
\begin{equation}
\Gamma(x,t)-\Gamma_{0}=\tilde{\Gamma}(t)\cos kx
\end{equation}
 and the initial condition $\Gamma(x,t=0)=\Gamma_{0}\ll\Gamma_{\mathrm{cmc}}$ where $\Gamma_\text{cmc}$ is the critical micelle concentration for surfactant, Eq.\,(\ref{eq:SURF-TRANS}) reduces to 
\begin{equation}
\frac{\partial\tilde{\Gamma}}{\partial t}+k^{2}D\tilde{\Gamma}=-\Gamma_{0}\frac{\partial u}{\partial x},\label{eq:surfactant}
\end{equation}
 where 
\begin{equation}
\frac{\partial u}{\partial x}=-\frac{\mathrm{d}a}{\mathrm{d}t}k+\frac{2\nu k^3}{\omega}\left(\frac{\alpha}{\mu}\tilde{\Gamma}+\frac{\mathrm{d}a}{\mathrm{d}t}\right)*\mathcal{F}.
\end{equation}

The resulting equation of motion is given by the simultaneous integro-differential
equation 
\begin{alignat}{1}
\ddot{a}+2\epsilon\dot{a}+a & =\epsilon\tilde{\Omega}(0,t)-2\epsilon^{2}\tilde{\Omega}(0,\tau)*\mathcal{F}(\tau),\label{eq:AMPLITUDE-EQN}\\
\dot{\tilde{\Gamma}}+\zeta\tilde{\Gamma} & =\delta\dot{a}+\epsilon\tilde{\Omega}(0,\tau)*\mathcal{F}(\tau),\label{eq:SURFACTANT-EQN}
\end{alignat}
where $\delta=a_{0}k,$ $\zeta=Dk^{2}/\omega$ for surface diffusivity
coefficient $D$.

Let $F(s)=\mathcal{L}[a](s)$, $G(s)=\mathcal{L}[\tilde{\Gamma}](s)$
be the Laplace transforms of $a(\tau)$ and $\tilde{\Gamma}(\tau)$,
respectively; the function $\hat{\Pi}_{\epsilon}\equiv sF(s)-a_{0}=\hat{\Pi}_{\epsilon}(s+\epsilon)=\mathrm{P}(s^{1/2})/\mathrm{Q}(s^{1/2})$
is given by 

\begin{widetext}
\begin{equation}
\hat{\Pi}_{\epsilon}=\frac{(u_{0}s-a_{0})\left\{ s(s+\zeta)+2\beta\epsilon^{1/2}[(s+\epsilon)^{1/2}-\epsilon^{1/2}]\right\} +2\epsilon\beta s(s-2\epsilon^{1/2}[(s+\epsilon)^{1/2}-\epsilon^{1/2}])}{s\epsilon^{1/2}\left\{ s(s+\zeta)+2\beta\epsilon^{1/2}[(s+\epsilon)^{1/2}-\epsilon^{1/2}]\right\} +2\epsilon\beta(s-2\epsilon^{1/2}[(s+\epsilon)^{1/2}-\epsilon^{1/2}])(\delta s-2\epsilon^{1/2}[(s+\epsilon)^{1/2}-\epsilon^{1/2}])},
\end{equation}

\end{widetext}

Let $\sigma_{i}^{(n)}$ be the $n$-th order cyclic polynomial given
by 
\begin{equation}
\sigma_{j}^{(n)}=\prod_{i=1}^{n-1}(z_{j+i\,\mathrm{mod}(n)}-z_{j}).
\end{equation}
Decomposing $\hat{\Pi}_{\epsilon}$ into partial fractions yields
\begin{equation}
\hat{\Pi}_{\epsilon}(s+\epsilon)=\sum_{i=1}^{8}\frac{c_{i}}{s^{1/2}+z_{i}}\label{eq:PI}
\end{equation}
where $-z_{i}$ are the roots of the polynomial $\mathrm{Q}(s^{1/2})$
and by comparison with Lagrange interpolation, the coefficients $c_{i}$
are given by the expression 
\begin{equation}
c_{i}=\frac{\mathrm{P}(-z_{i})}{\sigma_{i}^{(8)}(-z_{i})}.
\end{equation}
Taking the inverse Laplace transform of Eq. (\ref{eq:PI}) gives 
\begin{alignat}{1}
\Pi_{\epsilon}(\tau) & =-\sum_{i=1}^{8}\frac{\mathrm{P}(-z_{i})}{\sigma_{i}^{(8)}}z_{i}\mathrm{e}^{z_{i}^{2}\tau}\mathrm{erfc}(z_{i}\tau^{1/2}),
\end{alignat}
where we have used the fact (shown in Appendix A) that 
\begin{equation}
\deg\mathrm{Q}-\deg\mathrm{P}=2
\end{equation}
which implies $\mathrm{Z}(n,0)=0$, where 
\begin{equation}
\mathrm{Z}(n,j)=\sum_{i=1}^{n}\frac{\mathrm{P}(-z_{i})}{\sigma_{i}^{(n)}}(-z_{i})^{j}.
\end{equation}
Hence the amplitude is
\begin{alignat}{1}
a(\tau) & -a_{0}=\sum_{i=1}^{8}\frac{z_{i}}{\sigma_{i}^{(8)}}\mathrm{P}(-z_{i})\varphi(z_{i},\tau;\epsilon),\label{eq:CONVEC-SOL}
\end{alignat}
where $\varphi=\varphi(z_{i},\tau;\epsilon)$ is given by 
\begin{equation}
\varphi(z_{i},\tau;\epsilon)=\frac{\mathrm{e}^{(z_{i}^{2}-\epsilon)\tau}\mathrm{erfc}(z_{i}\tau^{1/2})+z_{i}\epsilon^{-1/2}\mathrm{erf}\sqrt{\epsilon\tau}-1}{z_{i}^{2}-\epsilon}.
\end{equation}

In the next section, we derive the explicit form of the amplitude solution in the special case near the critical wavelength where convective effects are neglected. 
}
\section{The exact diffusive-Marangoni Amplitude Solution}
Consider the solution
\begin{equation}
\Gamma_0-\Gamma(x,t)=\Gamma_{0}\mathrm{e}^{-k^2 Dt}\cos kx,\label{eq:ansatz}
\end{equation}
which satisfies the diffusively-dominant surfactant transport equation on the left-hand side of Eq.$\,$(\ref{eq:surfactant}) subjected to the augmented initial condition
\begin{equation}
	\Gamma(x,0)=\Gamma_0(1-\cos kx).
\end{equation}
This yields the interfacial vorticity 
\begin{equation}
\omega_{z}(x,f,t)=-2k\left(\frac{\alpha\Gamma_{0}}{\mu}\mathrm{e}^{-k^2 Dt}+\frac{\mathrm{d}a}{\mathrm{d}t}\right)\sin kx
\end{equation}
and so the boundary and initial conditions to Eq.\,(\ref{eq:Omega}) become
\begin{alignat}{1}
\Omega(0,t) & =-2k\left(\frac{\alpha\Gamma_{0}}{\mu}\mathrm{e}^{-k^2 Dt}+\frac{\mathrm{d}a}{\mathrm{d}t}\right),\\
\Omega(y,0) & =0.
\end{alignat}

Reverting to non-dimensional variables, taking the Laplace transform of Eq.\,(\ref{eq:EoM}) gives 
\begin{equation}
\beta^{-1}[F(s)-F_{\nu}(s)] =\frac{(s^{2}+2\epsilon s+1)\hat{\chi}(s+\epsilon)-1}{s(s+\zeta)},\label{eq:EoMLap}
\end{equation}
where 
\begin{equation}
	\hat{\chi}(s)=\left[s^{2}+2\epsilon s-4\epsilon^{3/2}s^{1/2}+\epsilon^{2}+1\right]^{-1},
\end{equation}
and $F_\nu(s)$ is the Laplace transform of the solution of the viscous amplitude equation in Eq.$\,$(\ref{eq:pEoM}) which satisfies the equation 
\begin{equation}
	sF_{\nu}(s)=a_{0}+(u_{0}s-a_{0})\hat{\chi}(s+\epsilon).
\end{equation}
Evaluating term-wise, the inverse Laplace transform of  $\beta^{-1}[F(s)-F_{\nu}(s)]$ is given by 
\begin{alignat}{1}
 & \beta^{-1}\mathcal{L}^{-1}[F(s)-F_{\nu}(s)](\tau)\\
=\ & \frac{\mathrm{e}^{-\zeta\tau}-1}{\zeta}-\sum_{i=1}^{4}\frac{z_{i}}{\sigma_{i}^{(3)}}\mathrm{e}^{(z_{i}^{2}-\epsilon)\tau}\mathrm{erfc}(z_{i}\sqrt{\tau})\nonumber \\
-\ & \sum_{i=1}^{4}\frac{z_{i}}{\sigma_{i}}\left[\frac{1}{\zeta}\varphi(z_{i},\tau;\epsilon)-\mathrm{e}^{-\zeta\tau}\times\right.\nonumber \\
 & \qquad \qquad\left.\left(2\epsilon-\zeta-\frac{1}{\zeta}\right)\varphi(z_{i},\tau;\epsilon-\zeta)\right],\label{eq:termwise}
\end{alignat}
where $z_{i}$ are the roots of the polynomial 
\begin{equation}
P(z)=z^{4}+2\epsilon z^{2}+4\epsilon^{3/2}z+\epsilon^{2}+1.
\end{equation}
To simplify Eq.\,(\ref{eq:termwise}), consider the sum 
\begin{equation}
Z(\epsilon;k)=\sum_{i=1}^{4}\frac{z_{i}^{k}}{\sigma_{i}^{(3)}(z_{i}^{2}-\epsilon)},
\end{equation}
then \cite{Prosperetti1976} gives the relations
\begin{alignat}{1}
Z(\epsilon;1) & =\frac{1+4\epsilon^{2}}{1+8\epsilon^{2}},\\
\epsilon^{-\frac{1}{2}}Z(\epsilon;2) & =-\frac{4\epsilon^{2}}{1+8\epsilon^{2}}.
\end{alignat}
Similarly, we have the relations (derivation in the Appendix B)

\begin{alignat}{1}
Z(\epsilon-\zeta;1) & =\frac{p_{-}+p_{+}}{2},\label{eq:Z1}\\
(\epsilon-\zeta)^{-\frac{1}{2}}Z(\epsilon;2) & =\frac{p_{-}-p_{+}}{2},\label{eq:Z2}
\end{alignat}
where 
\begin{equation}
	p_{\pm}=P(\pm(\epsilon-\zeta)^{\frac{1}{2}}).
\end{equation} 

Therefore, the solution to the full amplitude equation in Eq.$\,$(\ref{eq:EoM}) is given by 
\begin{alignat}{1}
a(\tau) & =a_{\nu}(\tau)-\frac{4\epsilon^{2}\beta}{(8\epsilon^{2}+1)\zeta}\mathrm{erfc}{\sqrt{\epsilon\tau}}\nonumber \\
 & +\left[\frac{1}{\zeta}+\left(2\epsilon-\zeta-\frac{1}{\zeta}\right)p_{+}\right]\beta\mathrm{e}^{-\zeta\tau}\\
 & +\frac{p_{+}-p_{-}}{2}\left(2\epsilon-\zeta-\frac{1}{\zeta}\right)\beta\mathrm{e}^{-\zeta\tau}\mathrm{erfc}\sqrt{(\epsilon-\zeta)\tau}\nonumber \\
 & -\sum_{i=1}^{4}\frac{z_{i}}{\sigma_{i}^{(3)}}\beta\mathrm{e}^{(z_{i}^{2}-\epsilon)\tau}\mathrm{erfc}(z_{i}\sqrt{\tau})\times\nonumber \\
 & \qquad\qquad\qquad\left(1+\frac{1}{\zeta(z_{i}^{2}-\epsilon)}+\frac{2\epsilon-\zeta-\frac{1}{\zeta}}{z_{i}^{2}-\epsilon+\zeta}\right),\label{eq:SOL}
\end{alignat}
where
\begin{alignat}{1}
a_{\nu}(\tau) & =\frac{4a_{0}\epsilon^{2}}{8\epsilon^{2}+1}\mathrm{erfc}\sqrt{\epsilon\tau}\nonumber \\
 & +\sum_{i=1}^{4}\frac{z_{i}}{\sigma_{i}^{(3)}}\left(\frac{a_{0}}{z_{i}^{2}-\epsilon}-u_{0}\right)\mathrm{e}^{(z_{i}^{2}-\epsilon)\tau}\mathrm{erfc}(z_{i}\sqrt{\tau})\label{eq:VISCOUS-SOL}
\end{alignat}
is the solution to the viscous amplitude equation \cite{Prosperetti1976} in Eq.$\,$(\ref{eq:pEoM}).

\changesss{
In the region near the critical wavelength $\lambda_{c}^{(0)},$ we
note that Marangoni diffusion is significant compared to Marangoni convection for $\mathrm{Sc}\ll 10^4$ in the regions near the critical wavelength.
To examine this in more detail, we define the Marangoni criticality coefficient $\varsigma$
by 
\begin{equation}
\varsigma=\frac{\zeta^{(1)}-\zeta^{(0)}}{\zeta^{(2)}-\zeta^{(0)}},\label{eq:varsigma}
\end{equation}
where $\zeta^{(0)}$ denotes the numerically calculated damping ratio of the capillary wave in the absence of Marangoni effect and $\zeta^{(1)},\zeta^{(2)}$ are the numerical damping ratios (1) with diffusional Marangoni effect and (2) with full
convective-diffusional Marangoni effect, respectively. At $\varsigma=\frac{1}{2}$, the effects of Marangoni convection contributes equally towards the global damping ratio as the Marangoni diffusion effects. Consequently, in the region
where $\varsigma>\frac{1}{2}$ the diffusional Marangoni effect
contributes more towards the overall damping ratio than the convectional
Marangoni effect and vice versa for $\varsigma<\frac{1}{2}$. Furthermore, the regions where $\varsigma > 0.95$
are defined as the \textit{Marangoni diffusion regime} where Marangoni diffusion dominates compared with the convection effects and is responsible for more than 95\% of the total Marangoni effect. Similarly, the region for which $\varsigma < 0.05$ denotes the \text{Marangoni convection regime}, which is not specifically investigated in this study. 

Under room temperature and pressure (rtp) with parameters density $\rho=10^{3}\mathrm{kgm}^{-3}$,
surface tension $\sigma=7.2\times10^{-2}\mathrm{Nm}^{-1}$ and viscosity
$\mu=10^{-3}\mathrm{Pa\,s}$, we plot $\varsigma$ for water
with respect to the non-dimensional wavelength $\lambda/\lambda_{c}^{(0)}$
normalised using the critical damping wavelength $\lambda_{c}^{(0)}$ in the absence of Marangoni effect, for
Schmidt number $\mathrm{Sc}=100$ in Fig.\,1(a) and $\mathrm{Sc}=500$
in Fig.\,1(b) with various Marangoni numbers $\mathrm{Ma}>500$. We notice that in both
cases, the diffusional effects as compared to the convection effects remain significant for roughly 2.5 critical wavelengths and dominate near the critical wavelength for $\mathrm{Sc}\sim O(10^2)$ and
$\mathrm{Ma}\sim O(10^2)$ as shown in Fig.\,1(a); it is only when we increase $\mathrm{Sc}$
that the diffusional Marangoni effect noticeably lessens near $\lambda_{c}^{(0)}$. This dominance of Marangoni diffusion over convection breaks down for $\mathrm{Sc}\gg O(10^2)$ and for wavelengths $\lambda\gg 2.5\lambda_c^{(0)}$, where Marangoni convection rapidly becomes more significant than diffusional effects. 
In the next section, we shall focus on the diffusion solution in Eq.\,(\ref{eq:SOL}) to analyse how the critical wavelength evolves
for increasing $\mathrm{Ma}$ within the Marangoni diffusion regime. 

\begin{figure}[htp]
\subfloat[]{
\includegraphics[width=4cm]{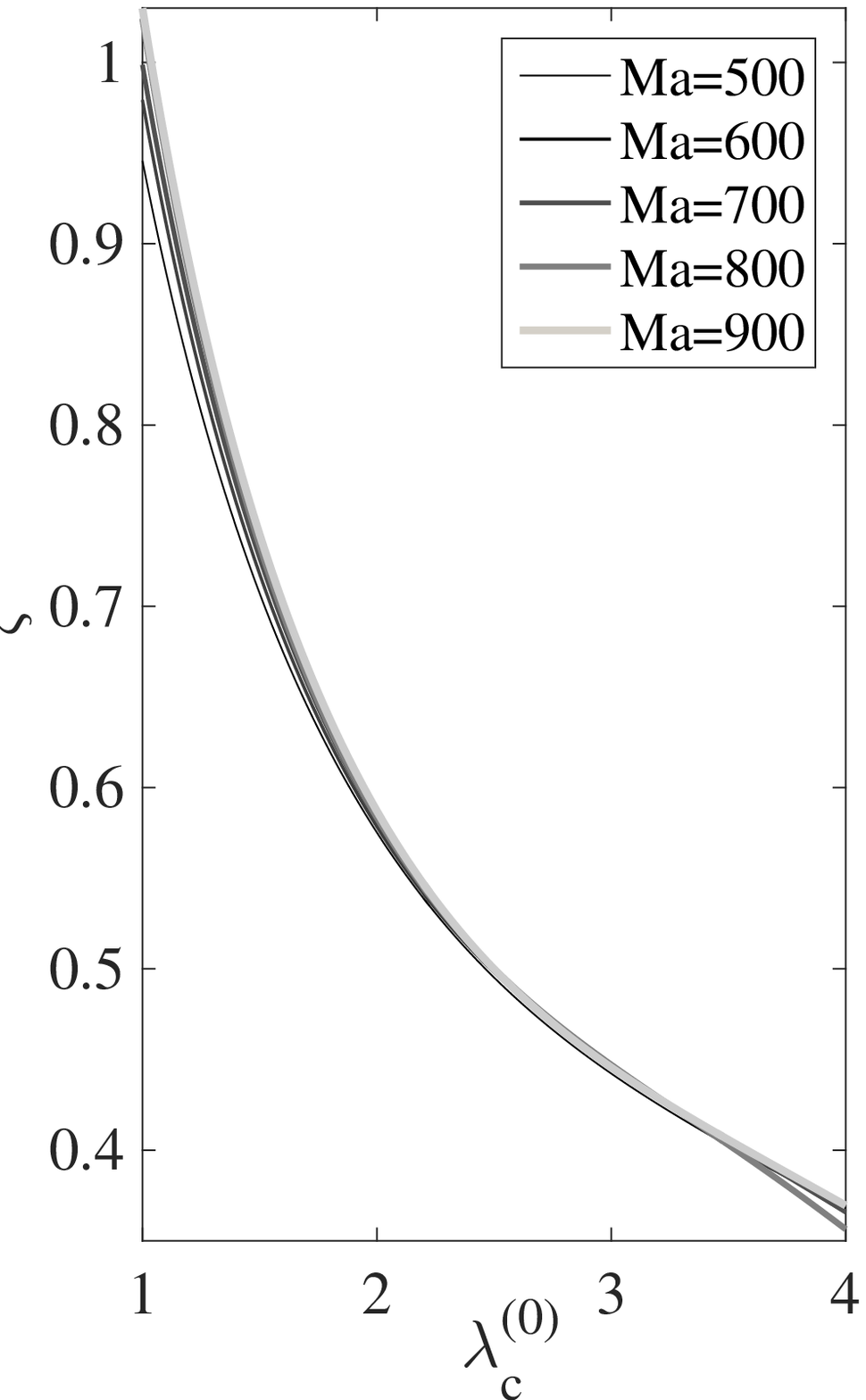}
} 
\subfloat[]{
\includegraphics[width=4cm]{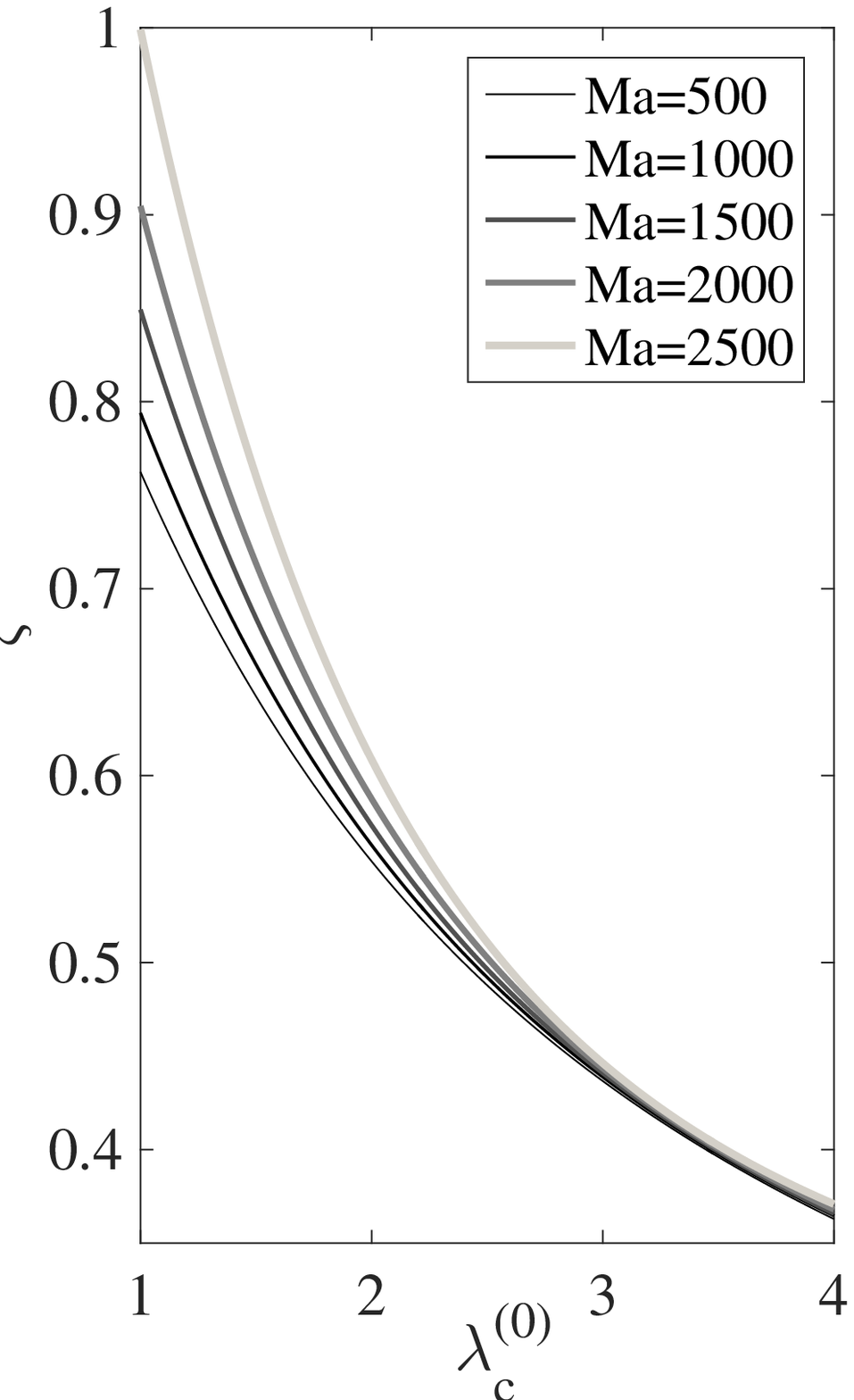}
} 

\caption{Comparison of the numerical coefficient $\varsigma$ defined in Eq.\,(\ref{eq:varsigma}) for (a) $\text{Sc}=100$ and (b) $\text{Sc}=500$ for the water interface at rtp. Marangoni convection dominates Marangoni diffusion in the region   $\varsigma<\frac{1}{2}$ and vice versa for $\varsigma<\frac{1}{2}$.}
\end{figure}
}

\section{Results and discussion}

Interactions of the viscosity and the Marangoni effect within the
fluid yield different consequences on the interface. While
both effects act as damping mechanisms to the surface waves \cite{Levich1969};
increasing the viscosity retards the rate at which vorticity generated
at the boundary enters the bulk \cite{Batchelor2000,Lamb1932}, whereas
increasing the concentration of surface-active substance \changes{endows} the interface with an effective surface shear and dilatational elasticity that works
to suppress surface motion \cite{Cantat2013,Levich1969}. 

\changess{We explore this viscosity-Marangoni interaction in the free-oscillation (i.e. $u_0=0$) solution in Eq.\,(\ref{eq:SOL}) by plotting the amplitude function $a(\tau)$ in Fig.\,2 for various Marangoni numbers and two viscosity values. Starting from the critically damped region with non-dimensional wavelength $\lambda/\lambda_\text{c}^{(0)}=1.5543$ in Fig.\,2(a) with $\mathrm{Sc}=20$; where $\lambda_\text{c}^{(0)}$ is the critical wavelength in absence of a surfactant solution; we observe that the addition of the Marangoni effect increases the damping of the capillary wave and the wave transitions from the underdamped to the overdamped as a result. For increasing Marangoni number, Fig.\,2(b) shows the evolution of the Marangoni correction given by $a_\text{M}(\tau)=a(\tau)-a_\nu(\tau)$, where $a_\nu(\tau)$ is the viscous solution in Eq.\,(\ref{eq:VISCOUS-SOL}). We note that the maximum of $a_\text{M}(\tau)$ is obtained near $\tau_c=\tau_c(\epsilon,\zeta)$ for all values of $\text{Ma}$ before decaying exponentially to zero. 

Furthermore, we increase the viscosity from $\mu_0=10^{-3}\text{Pa\,s}$ in Fig.\,2(a-b) to $1.3\mu_0$ in Fig.\,2(c-d) where the capillary wave is in the overdamped region where $\lambda/\lambda_\text{c}^{(0)}=0.5718$. From Fig.\,2(c), the increase in viscosity has three direct consequences; firstly the critical wavelength $\lambda_\text{c}^{(0)}$ increases, secondly, the Marangoni correction $a_\text{M}(\tau)$ becomes itself more damped for $\tau>\tau_c$ and finally, the time $\tau_c$ for which $a_\text{M}(\tau)$ obtains its maximum is now a weakly increasing function in $\text{Ma}$. 
}

\begin{figure}[htp]
\subfloat[]{
\includegraphics[width=8.6cm]{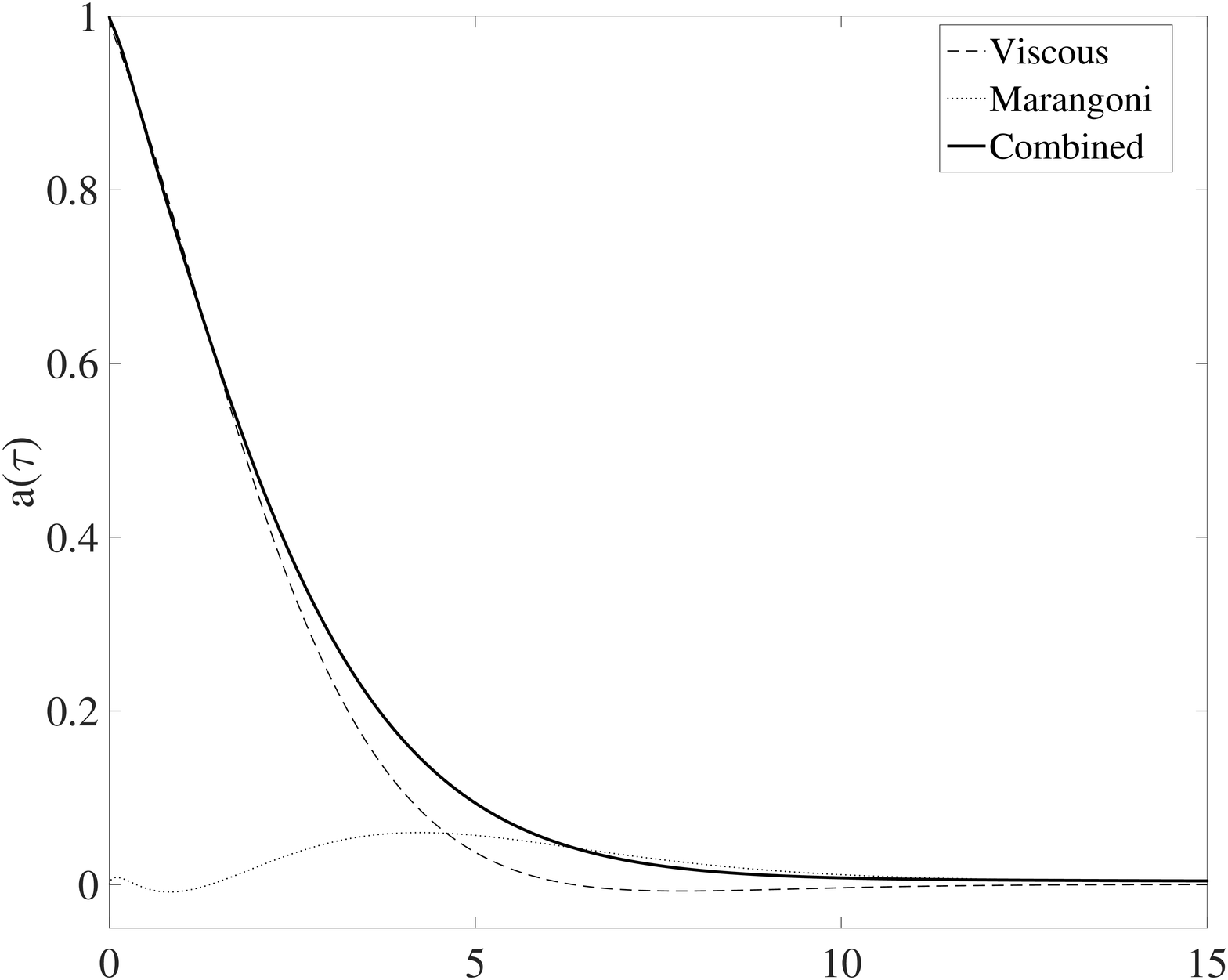}
}

\subfloat[]{
\includegraphics[width=4.3cm]{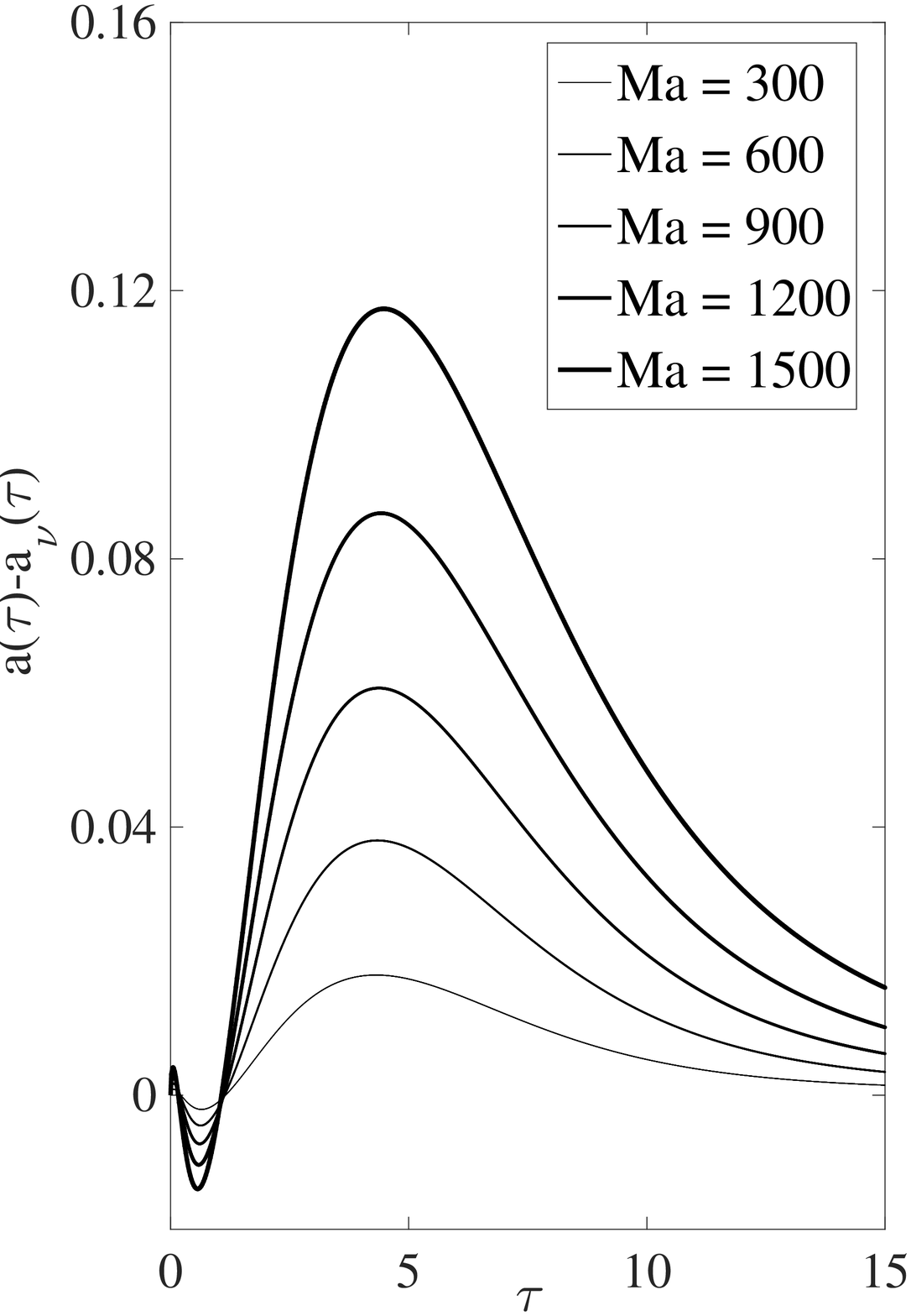}
}
\subfloat[]{
\includegraphics[width=4.3cm]{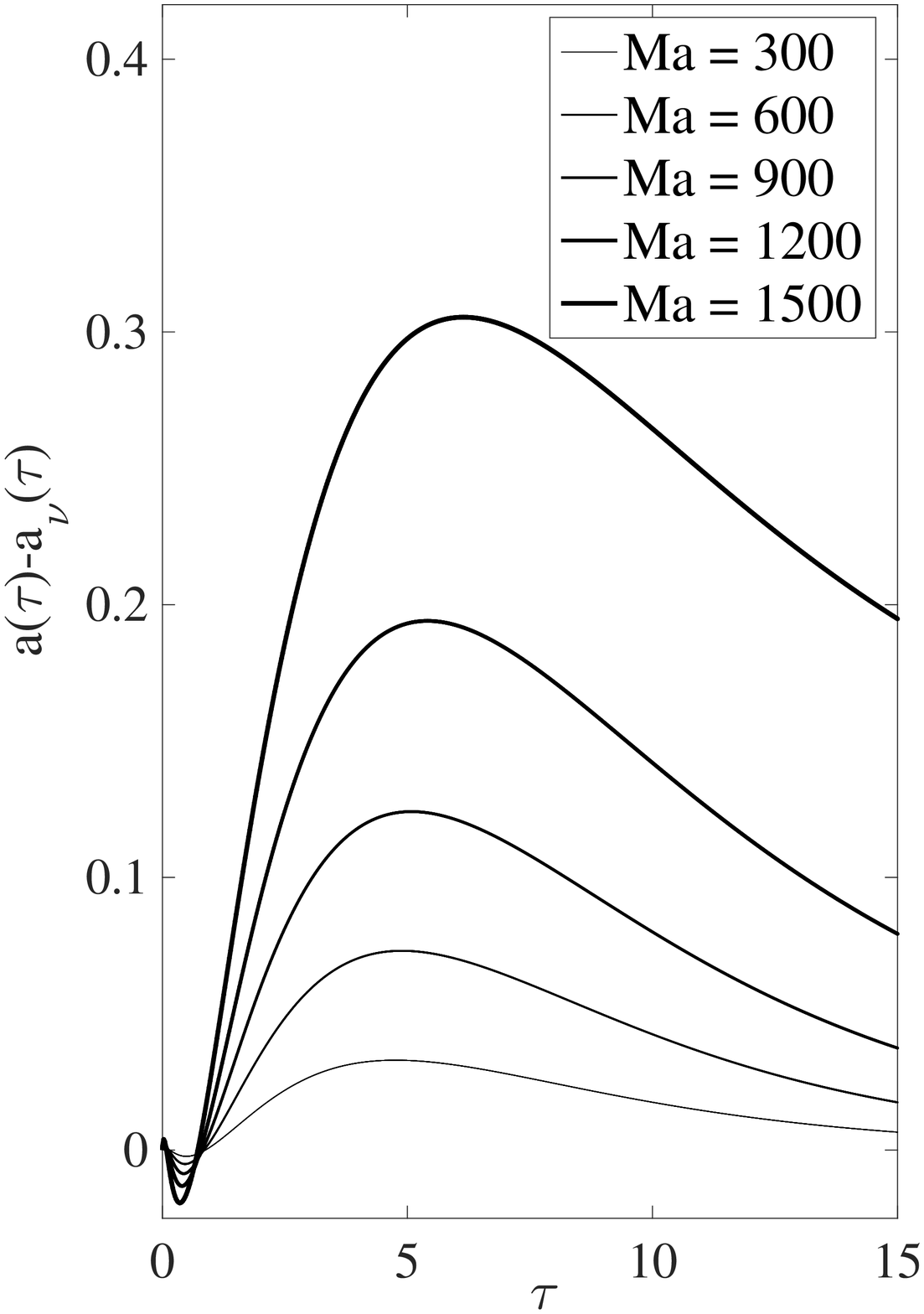}
} 

\subfloat[]{
\includegraphics[width=8.6cm]{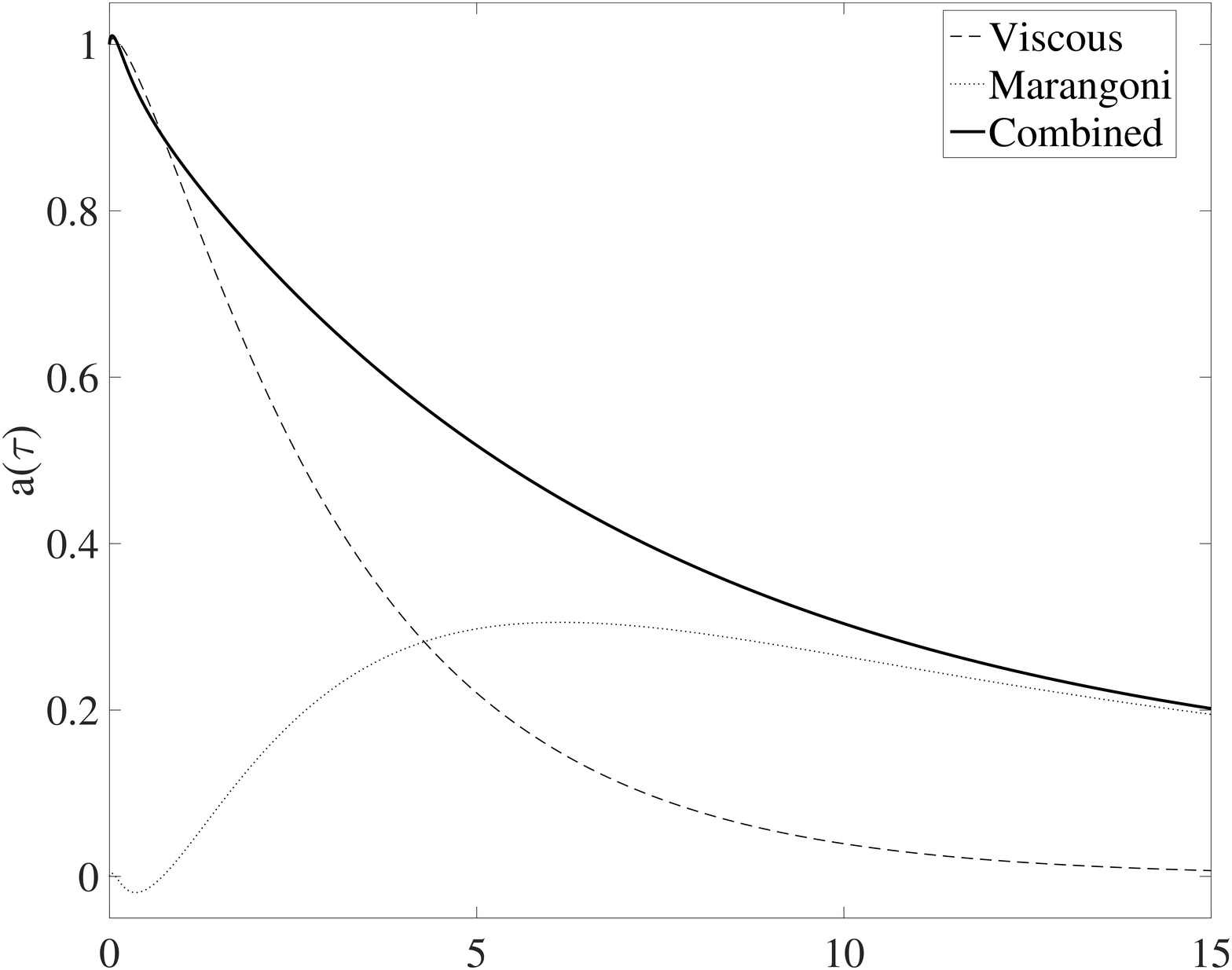}
} 

\caption{The free-oscillation solution in Eq.\,(\ref{eq:SOL}), for $\mathrm{Sc}=20$ and $\mathrm{Ma}=1500$, as a combination of Marangoni (dotted) and the viscous (dashed) contributions in (a) and (d) with $\bar{\lambda}\equiv\lambda/\lambda_\text{c}^{(0)}=1.5543$ and $0.5718$, respectively. In (b) and (c), the Marangoni correction $a(\tau)-a_\nu(\tau)$ for increasing Ma with $\bar{\lambda}=1.5543$ and $\bar{\lambda}=0.5718$, respectively; for the critically damped to overdamped regions in (a,b) where $\mu\equiv \mu_0=10^{-3}\mathrm{Pa\,s}$ and the overdamped region in (c,d) with $\mu=1.3\mu_0$.}
\end{figure}
\changess{It remains to show how the critical wavelength evolves for increasing
concentrations of surfactant under different $\mathrm{Sc}.$ Consider
firstly the case with no Marangoni effect, for which, based on a simple
damped harmonic oscillator with spring constant $c$ and mass $m$
in a viscous fluid, the displacement $s$ is given by
\begin{equation}
\frac{\mathrm{d}^2s}{\mathrm{d}\tau^2}+2d\omega_{0}\frac{\mathrm{d}s}{\mathrm{d}\tau}+\omega_{0}^{2}s=\mathrm{f}(\tau)\label{eq:lineareqn}
\end{equation}
where $d=b/(2\sqrt{mc})$ is the damping ratio for viscous damping
coefficient $b$, $\omega_{0}=\sqrt{c/m}$ is the undamped frequency and
$\mathrm{f}(\tau)=0$. Equating the spring constant $c$ with the surface tension
$\sigma$, let $m=\rho/k^{3}$ and consider the viscous damping coefficient
be defined as $b=\mu L_\mu$ where $L_\mu=\sqrt{2\mu/\rho\omega_{0}}$ is
the viscous damping length, Denner \cite{Denner2016b} proposed the critical wavelength in absence of the Marangoni effect
$\lambda_\mathrm{c}^{(0)}$ as 
\begin{equation}
\lambda_\mathrm{c}^{(0)}=\frac{2^{1/3}\pi}{\Theta}l_\text{vc}, \label{eq:CRIT-THEORY}
\end{equation}
for a single fluid with a free surface, where $\Theta=1.0625$ is a constant and $l_\text{vc}=\mu^{2}/\rho\sigma$ is the viscocapillary length scale. It has been shown \cite{Denner2016c} recently that this definition of the critical wavelength $\lambda_\text{c}^{(0)}$ also holds for capillary waves (under constant surface tension) with a finite amplitude.  

Switching on the Marangoni effect, we have the linearised external
forcing term $\mathrm{f}(\tau)=-2\epsilon\beta\mathrm{e}^{-\zeta\tau}$
in Eq.\,(\ref{eq:lineareqn}) which solves to give 
\begin{equation}
	s=A_{0}\mathrm{e}^{-2\epsilon\tau}\cos(B\tau+\phi_{0})-C\mathrm{e}^{-\zeta\tau}\label{eq:linearsol}
\end{equation}
where $B=(1-d^{2})^{1/2}$, $C=2\epsilon\beta\omega_{0}/(\zeta^{2}-2\zeta d\omega_{0}+\omega_{0}^{2})$ and $A_{0}=a_{0}/\cos\phi_{0}$ for phase angle $\phi_{0}$ satisfying $-B\omega_{0}\tan\phi_{0}=(u_{0}/a_{0})+2\epsilon$. Note that the ratio $2\epsilon/\zeta=2\mathrm{Sc}$ of the exponentials
in the linearised solution Eq.$\,$(\ref{eq:linearsol}) determines whether the motion favours
the sinusoidal viscous term or the second term  which arises from
the Marangoni correction. Moreover, the viscous damping factor $d$
deviates from the viscous case due to the presence of the surfactant
solution. We investigate this damping factor below.

Firstly, consider a range of $\mathrm{Sc}$ from orders $10^{1}$
to $10^{4}$ which is typical for chemical compounds in liquids under
room temperature \cite{Edwards1991}, we plot in Fig.$\,$3 the ratio of the surface tension difference and the initial surface tension $\sigma_0$ given by 
\begin{equation}
	\Sigma=\frac{\Delta\sigma}{\sigma_0}=\frac{\alpha\Delta\Gamma}{\sigma_0},
\end{equation}
against the ratio $\lambda_\mathrm{c}^{(\mathrm{Sc})}/\lambda_\mathrm{c}^{(0)}$,
where $\lambda_\mathrm{c}^{(\mathrm{Sc})}$ is the critical wavelength calculated
at $\mathrm{Sc}$. 

\begin{figure}[htp]
\includegraphics[width=7cm]{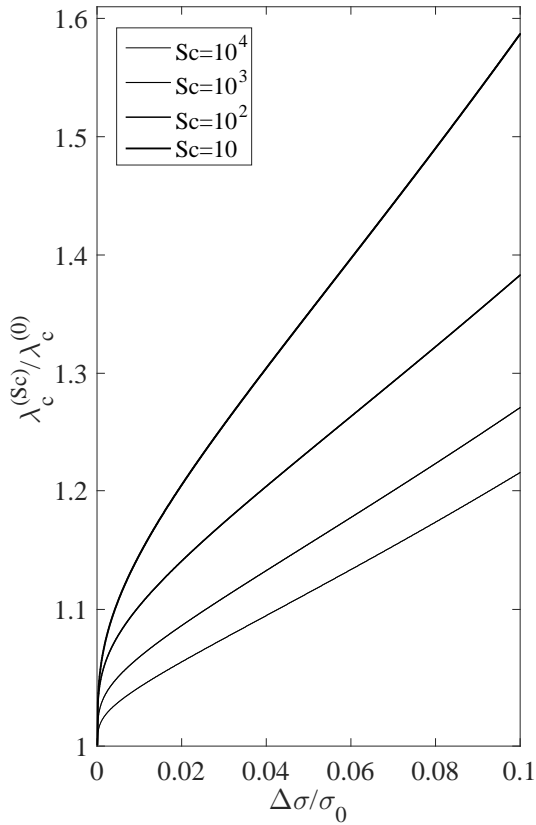}
\caption{Dimensionless critical wavelength $\lambda_\text{c}^{\text{Sc}}/\lambda_\text{c}^{(0)}$ (where $\lambda_\text{c}^{(0)}$ is the critical wavelength in absence of a surfactant solution) as a function of $\Sigma=\Delta\sigma/\sigma_0\in[0,\frac{1}{10}]$, the ratio of surfactant concentration difference and the initial surfactant concentration $\sigma_0$, for $\text{Sc}=10^1$ to $\text{Sc}=10^4$.}
\end{figure}

\begin{figure}[htp]
\subfloat[]{
\includegraphics[width=7cm]{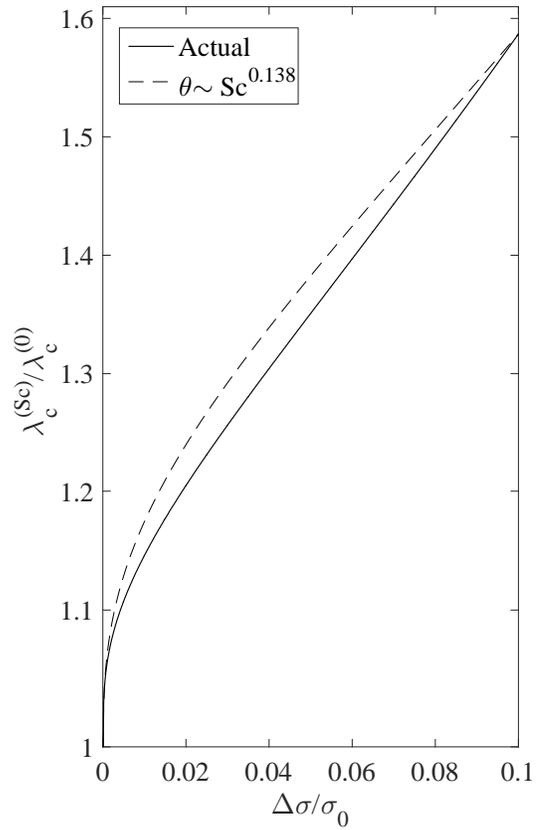}
}

\subfloat[]{
\includegraphics[width=7cm]{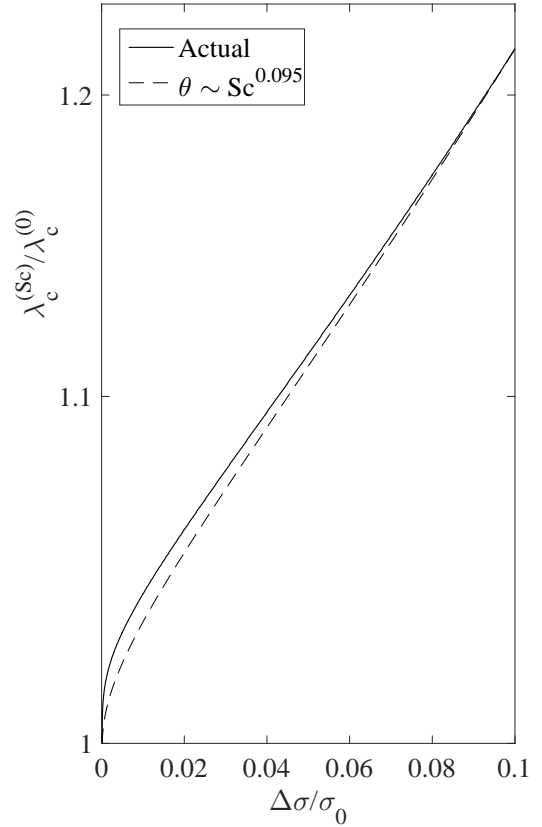}
}\caption{Dimensionless critical wavelength $\lambda_\text{c}^{\text{Sc}}/\lambda_\text{c}^{(0)}$ as a function of $\Sigma=\Delta\sigma/\sigma_0\in[0,\frac{1}{10}]$, the ratio of surfactant concentration difference and the initial surfactant concentration $\sigma_0$, for (a) upper bound for $\theta$ at $\text{Sc}=10^1$  and (b) lower bound for $\theta$ at $\mathrm{Sc=10^4}$.}
\end{figure}	

To derive an analytical scaling for the critical wavelength $\lambda_\mathrm{c}^{(\mathrm{Sc})}$
for different values of $\mathrm{Sc},$ we consider that the viscous
damping length $L_{\mu}$, which is the depth of penetration of the
vorticity generated by a capillary wave \cite{Landau1966,Denner2016b}, must
be augmented by a suitable diffusion damping length $L_{D}\sim\sqrt{D/\omega_{0}}$.
Dimensional analysis then gives 
\begin{equation}
L_{\mu,D}=\left(\frac{2\mu}{\rho\omega_{0}}\right)^{1/2}+\theta\left(\frac{2D\Sigma}{\omega_{0}}\right)^{1/2},
\end{equation}
where $\theta=\theta(\mathrm{Sc})$.

Consider the leading-order expansion 
\begin{equation}
	L_{\mu,D}^{2}=\frac{2\mu}{\rho \omega_0}\left[1+2\theta\sqrt{\frac{\Sigma}{\mathrm{Sc}}}+O\left(\frac{\Sigma}{\mathrm{Sc}}\right)\right],
\end{equation}
the critical damping relation 
\begin{equation}
d=\frac{b}{2\sqrt{mc}}=\frac{\mu L_{\mu,D}}{2\sqrt{\rho\sigma/k^{3}}}=1	
\end{equation}
yields the first-order expression for the critical wavelength with Marangoni effect  
\begin{equation}
	\lambda_\mathrm{c}^{(\mathrm{Sc})}(\Sigma) \sim 2^{1/3}\pi\, l_\text{vc}\left[1+\frac{4\theta}{3}\left(\frac{\Sigma}{\mathrm{Sc}}\right)^{1/3}\right].
\end{equation}

As $\mathrm{Sc}$ increases, it follows that $\theta$
decreases. Using a power-law form expression $\theta\sim\mathrm{Sc}^{\eta}$, we find numerically that $\eta$ increases monotonically in the range $\mathrm{Sc}\in[10,10^{4}]$ with approximate upper bound for $\mathrm{Sc}=10$ and lower bound for $\mathrm{Sc}=10^4$ shown in Fig.$\,$4(a) and (b), respectively. Taking an average value of the upper and lower bounds, we have the scaling
\begin{equation}
	\theta\sim\mathrm{Sc}^{1/9}.
\end{equation}
Under this scaling, the leading-order critical wavelength $\lambda_\mathrm{c}^{(\mathrm{Sc})}$
with diffusion-dominant Marangoni effect reduces to
\begin{alignat}{1}
\lambda_\mathrm{c}^{(\mathrm{Sc})}(\Sigma) & \simeq\frac{2^{1/3}\pi}{\Theta}l_\text{vc}\left[1+\frac{4}{3}\left(\frac{\Sigma}{\mathrm{Sc}^{2/3}}\right)^{1/3}\right],
\end{alignat}
which provides a first-order Marangoni correction term to the critical wavelength $\lambda_\text{c}^{(0)}=2^{1/3}\pi l_\text{vc}/\Theta$ obtained by Denner \cite{Denner2016b}.
}

\section{Conclusions}

To conclude, we derived a generalised integrodifferential initial
value problem for the wave amplitude of surface capillary waves in
the presence of the Marangoni effect which is solved exactly for a surfactant solution with concentration much less than
the critical micelle concentration (cmc) \changesss{under convective-diffusive surface transport. In particular, we investigated the diffusively-dominated region near the critical wavelength $\lambda_\text{c}^{(0)}$} and identified a first-order correction of $\lambda_\mathrm{c}^{(\text{Sc})}$ from $\lambda_\text{c}^{(0)}$  as a function of the Schmidt number $\mathrm{Sc}$ and $\Sigma=\Delta\sigma/\sigma_0$ the ratio of surface tension difference and the initial surface tension $\sigma_0$. This first-order correction provides an initial glimpse into the change of fundamental properties due to the Marangoni effect \changess{near the critical wavelength $\lambda_\text{c}^{(0)}$.}

\begin{acknowledgements}
The authors would like to acknowledge and thank the anonymous reviewers for the insightful comments and the constructive suggestions provided, which have lead to a more substantial discussion about the convective effects in the manuscript and a significant improvement of the quality of the final version of the manuscript. The authors would also like to acknowledge the financial support of the Shell University Technology Centre for fuels and lubricants and the Engineering and Physical Sciences Research Council (EPSRC) through grants EP/M021556/1 and EP/N025954/1. 		

\end{acknowledgements}
\changess{
\appendix

\section{$\deg\mathrm{Q}-\deg \mathrm{P}\geqslant2 \,\Rightarrow\, \mathrm{Z}(n,0)=0$}

Consider the rational expression 
\begin{alignat}{1}
\hat{\mathrm{f}}(s) & \equiv\frac{\mathrm{P}(s,m)}{\mathrm{Q}(s,n)}\label{eq:RATEXP}\\
 & =\frac{s^{m}+\varpi_{1}s^{m-1}+\cdots+\varpi_{m-1}s+\varpi_{m}}{s^{n}+\varsigma{}_{1}s^{n-1}+\cdots+\varsigma_{n-1}s+\varsigma_{n}}
\end{alignat}
where $\mathrm{P}(s,m)$ is a polynomial of order $m$ in $s$, 
\begin{equation}
\mathrm{Q}(s,n)=\prod_{i=1}^{n}(s-q_{i})
\end{equation}
 is a polynomial of order $n>m$ in $s$ with distinct roots $q_{i}$
and 
\begin{equation}
\sum_{1\leqslant i_{1}<i_{2}<\cdots<i_{k}\leqslant n}q_{i_{1}}q_{i_{2}}\cdots q_{i_{k}}=(-1)^{k}\varsigma_{n-k}.
\end{equation}

Rewriting $\hat{\mathrm{f}}(s)$ using a partial fraction decomposition,
we have 
\begin{equation}
\hat{\mathrm{f}}(s)=\sum_{i=1}^{n}\frac{\mathrm{P}(q_{i})}{\mathrm{Q}'(q_{i})}\frac{1}{s-q_{i}}
\end{equation}
and taking an inverse Laplace transform gives 
\begin{alignat}{1}
\mathrm{f}(t) & =\sum_{i=1}^{n}\frac{\mathrm{P}(q_{i})}{\sigma_{i}^{(n)}(q_{i})}\mathrm{e}^{-q_{i}t}\\
 & =\sum_{j=0}^{\infty}(-1)^{j}\mathrm{Z}(n,j)\frac{t^{j}}{j!},
\end{alignat}
where $q_{i}$ are roots of the polynomial $\mathrm{Q}(s,n)$ and
\begin{equation}
\mathrm{Z}(n,j)=\sum_{k=1}^{n}\frac{\mathrm{P}(q_{j})}{\sigma_{j}^{(n)}(q_{j})}q_{k}^{j}.\label{eq:Z(NJ)}
\end{equation}
Expansion of Eq.$\,$(\ref{eq:RATEXP}) for large $s$ and inversion
term-wise gives 
\begin{equation}
\mathrm{f}(t)\sim\frac{t^{n-m-1}}{(n-m-1)!}+\frac{(\varpi_{1}-\varsigma_{1}\varpi_{m})t^{n-m}}{(n-m)!}+O(t^{n-m+1}).
\end{equation}
Comparing with Eq.$\,$(\ref{eq:Z(NJ)}) shows that $\mathrm{Z}(n,j)=0$
 if $0\leqslant j\leqslant n-m-2$, i.e. we need
\begin{equation}
\deg\mathrm{Q}-\deg\mathrm{P}\geqslant2.
\end{equation}

\section{$Z(\epsilon-\zeta;k)$}	In the exact solution in section V, we have simplified the solution
using Eq.\,(\ref{eq:Z1}) and (\ref{eq:Z2}), modifying a similar approach to \cite{Prosperetti1976}.
We consider distinct $c_{1},\ldots,c_{n}\in\mathbb{C}$ and define 
\begin{alignat}{1}
p_{n,j} & =\prod_{i=1}^{n-1}\left(c_{j+i\,\mathrm{mod}(n)}-c_{j}\right)\\
S_{n}(k) & =\sum_{j=1}^{n}\frac{c_{j}^{k}}{p_{n,j}},\quad k\in\mathbb{N},
\end{alignat}
then Prosperetti \cite{Prosperetti1976} showed that 
\begin{equation}
S_{n}(k)=\begin{cases}
0 & k=0,1,2,\ldots,n-2\\
(-1)^{n-1} & k=n-1\\
(-1)^{n-1}\sum_{i=1}^{n}c_{i} & k=n.
\end{cases}
\end{equation}
Let
\begin{alignat}{1}
Z(\epsilon-\zeta;k) & =\sum_{i=1}^{4}\frac{z_{i}^{k}}{\sigma_{i}(z_{i}^{2}-\epsilon+\zeta)},
\end{alignat}
$c_{i}=z_{i}$ for $1\leqslant i\leqslant4$ and define
\begin{alignat}{1}
c_5 & =(\epsilon-\zeta)^{1/2},\ \\
c_6 & =-(\epsilon-\zeta)^{1/2},
\end{alignat}
then we have 
\begin{alignat}{1}
p_{6,5} & =-2(\epsilon-\zeta)^{1/2}P[-(\epsilon-\zeta)^{1/2}],\ \\
p_{6,6} & =2(\epsilon-\zeta)^{1/2}P[(\epsilon-\zeta)^{1/2}]
\end{alignat}
where $P(z)$ is the polynomial given by 
\begin{equation}
	P(z)=z^{4}+2\epsilon z^{2}+4\epsilon^{3/2}z+1+\epsilon^{2}.
\end{equation}
Let $p_\pm=P(\pm(\epsilon-\zeta)^{1/2})$, then 
\begin{alignat}{1}
p_{6,5} & =-2(\epsilon-\zeta)^{1/2}/p_{-},\\
p_{6,6} & =2(\epsilon-\zeta)^{1/2}/p_{+}.
\end{alignat}
It follows that 
\begin{equation}
S_{6}(k)=Z(\epsilon-\zeta;k)+\frac{c_{5}^{k}}{p_{6,5}}+\frac{c_{6}^{k}}{p_{6,6}}=0,
\end{equation}
 and so 
\begin{alignat}{1}
Z(\epsilon-\zeta;1) & =-\left(\frac{c_{5}}{p_{6,5}}+\frac{c_{6}}{p_{6,6}}\right)\nonumber \\
 & =\frac{p_{-}+p_{+}}{2},\label{eq:Z1}
\end{alignat}
and 
\begin{alignat}{1}
Z(\epsilon-\zeta;2) & =-\left(\frac{c_{5}^{2}}{p_{6,5}}+\frac{c_{6}^{2}}{p_{6,6}}\right)\nonumber \\
 & =(\epsilon-\zeta)^{1/2}\left(\frac{p_{-}-p_{+}}{2}\right).\label{eq:Z2}
\end{alignat}

 }	
	
\bibliographystyle{apsrev4-1}
\bibliography{Bib.bib}

\begin{thebibliography}{30}%
\makeatletter
\providecommand \@ifxundefined [1]{%
 \@ifx{#1\undefined}
}%
\providecommand \@ifnum [1]{%
 \ifnum #1\expandafter \@firstoftwo
 \else \expandafter \@secondoftwo
 \fi
}%
\providecommand \@ifx [1]{%
 \ifx #1\expandafter \@firstoftwo
 \else \expandafter \@secondoftwo
 \fi
}%
\providecommand \natexlab [1]{#1}%
\providecommand \enquote  [1]{``#1''}%
\providecommand \bibnamefont  [1]{#1}%
\providecommand \bibfnamefont [1]{#1}%
\providecommand \citenamefont [1]{#1}%
\providecommand \href@noop [0]{\@secondoftwo}%
\providecommand \href [0]{\begingroup \@sanitize@url \@href}%
\providecommand \@href[1]{\@@startlink{#1}\@@href}%
\providecommand \@@href[1]{\endgroup#1\@@endlink}%
\providecommand \@sanitize@url [0]{\catcode `\\12\catcode `\$12\catcode
  `\&12\catcode `\#12\catcode `\^12\catcode `\_12\catcode `\%12\relax}%
\providecommand \@@startlink[1]{}%
\providecommand \@@endlink[0]{}%
\providecommand \url  [0]{\begingroup\@sanitize@url \@url }%
\providecommand \@url [1]{\endgroup\@href {#1}{\urlprefix }}%
\providecommand \urlprefix  [0]{URL }%
\providecommand \Eprint [0]{\href }%
\providecommand \doibase [0]{http://dx.doi.org/}%
\providecommand \selectlanguage [0]{\@gobble}%
\providecommand \bibinfo  [0]{\@secondoftwo}%
\providecommand \bibfield  [0]{\@secondoftwo}%
\providecommand \translation [1]{[#1]}%
\providecommand \BibitemOpen [0]{}%
\providecommand \bibitemStop [0]{}%
\providecommand \bibitemNoStop [0]{.\EOS\space}%
\providecommand \EOS [0]{\spacefactor3000\relax}%
\providecommand \BibitemShut  [1]{\csname bibitem#1\endcsname}%
\let\auto@bib@innerbib\@empty
\bibitem [{\citenamefont {Aarts}\ \emph {et~al.}(2004)\citenamefont {Aarts},
  \citenamefont {Schmidt},\ and\ \citenamefont {Lekkerkerker}}]{Aarts2004}%
  \BibitemOpen
  \bibfield  {author} {\bibinfo {author} {\bibfnamefont {D.~G.}\ \bibnamefont
  {Aarts}}, \bibinfo {author} {\bibfnamefont {M.}~\bibnamefont {Schmidt}}, \
  and\ \bibinfo {author} {\bibfnamefont {H.~N.}\ \bibnamefont {Lekkerkerker}},\
  }\href {\doibase 10.1126/science.1097116} {\bibfield  {journal} {\bibinfo
  {journal} {Science}\ }\textbf {\bibinfo {volume} {304}},\ \bibinfo {pages}
  {847} (\bibinfo {year} {2004})}\BibitemShut {NoStop}%
\bibitem [{\citenamefont {Mandelstam}(1913)}]{Mandelstam1913}%
  \BibitemOpen
  \bibfield  {author} {\bibinfo {author} {\bibfnamefont {L.}~\bibnamefont
  {Mandelstam}},\ }\href
  {http://onlinelibrary.wiley.com/doi/10.1002/andp.19133460808/abstract}
  {\bibfield  {journal} {\bibinfo  {journal} {Ann. Phys.}\ }\textbf {\bibinfo
  {volume} {346}},\ \bibinfo {pages} {609} (\bibinfo {year}
  {1913})}\BibitemShut {NoStop}%
\bibitem [{\citenamefont {Prosperetti}(1976)}]{Prosperetti1976}%
  \BibitemOpen
  \bibfield  {author} {\bibinfo {author} {\bibfnamefont {A.}~\bibnamefont
  {Prosperetti}},\ }\href {\doibase 10.1063/1.861446} {\bibfield  {journal}
  {\bibinfo  {journal} {Phys. Fluids}\ }\textbf {\bibinfo {volume} {19}},\
  \bibinfo {pages} {195} (\bibinfo {year} {1976})}\BibitemShut {NoStop}%
\bibitem [{\citenamefont {Lamb}(1932)}]{Lamb1932}%
  \BibitemOpen
  \bibfield  {author} {\bibinfo {author} {\bibfnamefont {S.~H.}\ \bibnamefont
  {Lamb}},\ }\href@noop {} {\emph {\bibinfo {title} {Hydrodynamics}}},\
  \bibinfo {edition} {6th}\ ed.\ (\bibinfo  {publisher} {Cambridge University
  Press},\ \bibinfo {year} {1932})\BibitemShut {NoStop}%
\bibitem [{\citenamefont {Scheludko}(1967)}]{Scheludko1967}%
  \BibitemOpen
  \bibfield  {author} {\bibinfo {author} {\bibfnamefont {A.}~\bibnamefont
  {Scheludko}},\ }\href {\doibase 10.1016/0001-8686(67)85001-2} {\bibfield
  {journal} {\bibinfo  {journal} {Adv. Colloid Interface Sci.}\ }\textbf
  {\bibinfo {volume} {1}},\ \bibinfo {pages} {391} (\bibinfo {year}
  {1967})}\BibitemShut {NoStop}%
\bibitem [{\citenamefont {Blanchette}\ and\ \citenamefont
  {Bigioni}(2006)}]{Blanchette2006}%
  \BibitemOpen
  \bibfield  {author} {\bibinfo {author} {\bibfnamefont {F.}~\bibnamefont
  {Blanchette}}\ and\ \bibinfo {author} {\bibfnamefont {T.~P.}\ \bibnamefont
  {Bigioni}},\ }\href {\doibase 10.1038/nphys268} {\bibfield  {journal}
  {\bibinfo  {journal} {Nat. Phys.}\ }\textbf {\bibinfo {volume} {2}},\
  \bibinfo {pages} {254} (\bibinfo {year} {2006})}\BibitemShut {NoStop}%
\bibitem [{\citenamefont {Sferrazza}\ \emph {et~al.}(1997)\citenamefont
  {Sferrazza}, \citenamefont {Xiao}, \citenamefont {Jones}, \citenamefont
  {Bucknall}, \citenamefont {Webster},\ and\ \citenamefont
  {Penfold}}]{Sferrazza1997}%
  \BibitemOpen
  \bibfield  {author} {\bibinfo {author} {\bibfnamefont {M.}~\bibnamefont
  {Sferrazza}}, \bibinfo {author} {\bibfnamefont {C.}~\bibnamefont {Xiao}},
  \bibinfo {author} {\bibfnamefont {R.~A.~L.}\ \bibnamefont {Jones}}, \bibinfo
  {author} {\bibfnamefont {D.~G.}\ \bibnamefont {Bucknall}}, \bibinfo {author}
  {\bibfnamefont {J.}~\bibnamefont {Webster}}, \ and\ \bibinfo {author}
  {\bibfnamefont {J.}~\bibnamefont {Penfold}},\ }\href {\doibase
  10.1103/PhysRevLett.78.3693} {\bibfield  {journal} {\bibinfo  {journal}
  {Phys. Rev. Lett.}\ }\textbf {\bibinfo {volume} {78}},\ \bibinfo {pages}
  {3693} (\bibinfo {year} {1997})}\BibitemShut {NoStop}%
\bibitem [{\citenamefont {Saye}\ and\ \citenamefont
  {Sethian}(2013)}]{Saye2013}%
  \BibitemOpen
  \bibfield  {author} {\bibinfo {author} {\bibfnamefont {R.}~\bibnamefont
  {Saye}}\ and\ \bibinfo {author} {\bibfnamefont {J.}~\bibnamefont {Sethian}},\
  }\href {\doibase 10.1126/science.1230623} {\bibfield  {journal} {\bibinfo
  {journal} {Science}\ }\textbf {\bibinfo {volume} {340}},\ \bibinfo {pages}
  {720} (\bibinfo {year} {2013})}\BibitemShut {NoStop}%
\bibitem [{\citenamefont {Papageorgiou}(1993)}]{Papageorgiou1993}%
  \BibitemOpen
  \bibfield  {author} {\bibinfo {author} {\bibfnamefont {D.~T.}\ \bibnamefont
  {Papageorgiou}},\ }\href {\doibase 10.1017/S002211209500382X} {\bibfield
  {journal} {\bibinfo  {journal} {J. Fluid Mech.}\ }\textbf {\bibinfo {volume}
  {301}},\ \bibinfo {pages} {109} (\bibinfo {year} {1993})}\BibitemShut
  {NoStop}%
\bibitem [{\citenamefont {Eggers}\ and\ \citenamefont
  {Villermaux}(2008)}]{Eggers2008}%
  \BibitemOpen
  \bibfield  {author} {\bibinfo {author} {\bibfnamefont {J.}~\bibnamefont
  {Eggers}}\ and\ \bibinfo {author} {\bibfnamefont {E.}~\bibnamefont
  {Villermaux}},\ }\href {\doibase 10.1088/0034-4885/71/3/036601} {\bibfield
  {journal} {\bibinfo  {journal} {Rep. Prog. Phys.}\ }\textbf {\bibinfo
  {volume} {71}},\ \bibinfo {pages} {036601} (\bibinfo {year}
  {2008})}\BibitemShut {NoStop}%
\bibitem [{\citenamefont {Hoepffner}\ and\ \citenamefont
  {Par{\'{e}}}(2013)}]{Hoepffner2013}%
  \BibitemOpen
  \bibfield  {author} {\bibinfo {author} {\bibfnamefont {J.}~\bibnamefont
  {Hoepffner}}\ and\ \bibinfo {author} {\bibfnamefont {G.}~\bibnamefont
  {Par{\'{e}}}},\ }\href {\doibase 10.1017/jfm.2013.472} {\bibfield  {journal}
  {\bibinfo  {journal} {J. Fluid Mech.}\ }\textbf {\bibinfo {volume} {734}},\
  \bibinfo {pages} {183} (\bibinfo {year} {2013})}\BibitemShut {NoStop}%
\bibitem [{\citenamefont {Lhuissier}\ \emph {et~al.}(2016)\citenamefont
  {Lhuissier}, \citenamefont {Brunet},\ and\ \citenamefont
  {Dorbolo}}]{Lhuissier2016}%
  \BibitemOpen
  \bibfield  {author} {\bibinfo {author} {\bibfnamefont {H.}~\bibnamefont
  {Lhuissier}}, \bibinfo {author} {\bibfnamefont {P.}~\bibnamefont {Brunet}}, \
  and\ \bibinfo {author} {\bibfnamefont {S.}~\bibnamefont {Dorbolo}},\ }\href
  {\doibase 10.1017/jfm.2016.241} {\bibfield  {journal} {\bibinfo  {journal}
  {J. Fluid Mech.}\ }\textbf {\bibinfo {volume} {795}},\ \bibinfo {pages} {784}
  (\bibinfo {year} {2016})}\BibitemShut {NoStop}%
\bibitem [{\citenamefont {Batchelor}\ \emph {et~al.}(2003)\citenamefont
  {Batchelor}, \citenamefont {Moffatt}, \citenamefont {Worster},\ and\
  \citenamefont {Osborn}}]{Batchelor2003}%
  \BibitemOpen
  \bibfield  {author} {\bibinfo {author} {\bibfnamefont {G.~K.}\ \bibnamefont
  {Batchelor}}, \bibinfo {author} {\bibfnamefont {H.~K.}\ \bibnamefont
  {Moffatt}}, \bibinfo {author} {\bibfnamefont {M.~G.}\ \bibnamefont
  {Worster}}, \ and\ \bibinfo {author} {\bibfnamefont {T.~R.}\ \bibnamefont
  {Osborn}},\ }\href {\doibase 10.1115/1.1603306} {\emph {\bibinfo {title}
  {Perspectives in Fluid Dynamics}}}\ (\bibinfo  {publisher} {CUP},\ \bibinfo
  {year} {2003})\BibitemShut {NoStop}%
\bibitem [{\citenamefont {Karakashev}\ and\ \citenamefont
  {Manev}(2014)}]{Karakashev2014}%
  \BibitemOpen
  \bibfield  {author} {\bibinfo {author} {\bibfnamefont {S.~I.}\ \bibnamefont
  {Karakashev}}\ and\ \bibinfo {author} {\bibfnamefont {E.~D.}\ \bibnamefont
  {Manev}},\ }\href {\doibase 10.1016/j.cis.2014.07.010} {\bibfield  {journal}
  {\bibinfo  {journal} {Adv. Colloid Interface Sci.}\ ,\ \bibinfo {pages} {1}}
  (\bibinfo {year} {2014})}\BibitemShut {NoStop}%
\bibitem [{\citenamefont {Sides}\ \emph {et~al.}(1999)\citenamefont {Sides},
  \citenamefont {Grest},\ and\ \citenamefont {Lacasse}}]{Sides1999}%
  \BibitemOpen
  \bibfield  {author} {\bibinfo {author} {\bibfnamefont {S.}~\bibnamefont
  {Sides}}, \bibinfo {author} {\bibfnamefont {G.}~\bibnamefont {Grest}}, \ and\
  \bibinfo {author} {\bibfnamefont {M.-D.}\ \bibnamefont {Lacasse}},\ }\href
  {\doibase 10.1103/PhysRevE.60.6708} {\bibfield  {journal} {\bibinfo
  {journal} {Phys. Rev. E}\ }\textbf {\bibinfo {volume} {60}},\ \bibinfo
  {pages} {6708} (\bibinfo {year} {1999})}\BibitemShut {NoStop}%
\bibitem [{\citenamefont {Chandrasekhar}(1961)}]{Chandrasekhar1981}%
  \BibitemOpen
  \bibfield  {author} {\bibinfo {author} {\bibfnamefont {S.}~\bibnamefont
  {Chandrasekhar}},\ }\href {\doibase 10.1017/S0022112062210592} {\emph
  {\bibinfo {title} {Hydrodynamic and hydromagnetic stability}}}\ (\bibinfo
  {publisher} {Dover Publications},\ \bibinfo {year} {1961})\BibitemShut
  {NoStop}%
\bibitem [{\citenamefont {Delgado-Buscalioni}\ \emph
  {et~al.}(2008)\citenamefont {Delgado-Buscalioni}, \citenamefont {Chacon},\
  and\ \citenamefont {Tarazona}}]{Delgado2008a}%
  \BibitemOpen
  \bibfield  {author} {\bibinfo {author} {\bibfnamefont {R.}~\bibnamefont
  {Delgado-Buscalioni}}, \bibinfo {author} {\bibfnamefont {E.}~\bibnamefont
  {Chacon}}, \ and\ \bibinfo {author} {\bibfnamefont {P.}~\bibnamefont
  {Tarazona}},\ }\href {\doibase 10.1103/PhysRevLett.101.106102} {\bibfield
  {journal} {\bibinfo  {journal} {Phys. Rev. Lett.}\ }\textbf {\bibinfo
  {volume} {101}},\ \bibinfo {pages} {106102} (\bibinfo {year}
  {2008})}\BibitemShut {NoStop}%
\bibitem [{\citenamefont {J{\"{a}}ckle}(1999)}]{Jackle1999}%
  \BibitemOpen
  \bibfield  {author} {\bibinfo {author} {\bibfnamefont {J.}~\bibnamefont
  {J{\"{a}}ckle}},\ }\href {\doibase 10.1088/0953-8984/10/32/004} {\bibfield
  {journal} {\bibinfo  {journal} {J. Phys. Condens. Matter}\ }\textbf {\bibinfo
  {volume} {10}},\ \bibinfo {pages} {7121} (\bibinfo {year}
  {1999})}\BibitemShut {NoStop}%
\bibitem [{\citenamefont {Denner}(2016)}]{Denner2016b}%
  \BibitemOpen
  \bibfield  {author} {\bibinfo {author} {\bibfnamefont {F.}~\bibnamefont
  {Denner}},\ }\href@noop {} {\bibfield  {journal} {\bibinfo  {journal}
  {Physical Review E}\ }\textbf {\bibinfo {volume} {94}},\ \bibinfo {pages}
  {023110} (\bibinfo {year} {2016})}\BibitemShut {NoStop}%
\bibitem [{\citenamefont {Yaminsky}\ \emph {et~al.}(2010)\citenamefont
  {Yaminsky}, \citenamefont {Ohnishi}, \citenamefont {Vogler},\ and\
  \citenamefont {Horn}}]{Yaminsky2010}%
  \BibitemOpen
  \bibfield  {author} {\bibinfo {author} {\bibfnamefont {V.~V.}\ \bibnamefont
  {Yaminsky}}, \bibinfo {author} {\bibfnamefont {S.}~\bibnamefont {Ohnishi}},
  \bibinfo {author} {\bibfnamefont {E.~A.}\ \bibnamefont {Vogler}}, \ and\
  \bibinfo {author} {\bibfnamefont {R.~G.}\ \bibnamefont {Horn}},\ }\href
  {\doibase 10.1021/la904481d} {\bibfield  {journal} {\bibinfo  {journal}
  {Langmuir}\ }\textbf {\bibinfo {volume} {26}},\ \bibinfo {pages} {8061}
  (\bibinfo {year} {2010})}\BibitemShut {NoStop}%
\bibitem [{\citenamefont {{Cantat I, Cohen-Addad S, Elias F, Graner F,
  H{\"{o}}hler R}}(2013)}]{Cantat2013}%
  \BibitemOpen
  \bibfield  {author} {\bibinfo {author} {\bibfnamefont {P.~O.}\ \bibnamefont
  {{Cantat I, Cohen-Addad S, Elias F, Graner F, H{\"{o}}hler R}}},\ }\href@noop
  {} {\emph {\bibinfo {title} {{Foams: Structure and Dynamics}}}}\ (\bibinfo
  {publisher} {Oxford University Press},\ \bibinfo {year} {2013})\BibitemShut
  {NoStop}%
\bibitem [{\citenamefont {Batchelor}(2000)}]{Batchelor2000}%
  \BibitemOpen
  \bibfield  {author} {\bibinfo {author} {\bibfnamefont {G.~K.}\ \bibnamefont
  {Batchelor}},\ }\href {\doibase 10.1063/1.3060769} {\emph {\bibinfo {title}
  {An Introduction to Fluid Dynamics}}}\ (\bibinfo  {publisher} {Cambridge
  University Press},\ \bibinfo {year} {2000})\BibitemShut {NoStop}%
\bibitem [{\citenamefont {Denner}\ \emph {et~al.}(2017)\citenamefont {Denner},
  \citenamefont {Par{\'{e}}},\ and\ \citenamefont {Zaleski}}]{Denner2016c}%
  \BibitemOpen
  \bibfield  {author} {\bibinfo {author} {\bibfnamefont {F.}~\bibnamefont
  {Denner}}, \bibinfo {author} {\bibfnamefont {G.}~\bibnamefont {Par{\'{e}}}},
  \ and\ \bibinfo {author} {\bibfnamefont {S.}~\bibnamefont {Zaleski}},\
  }\href@noop {} {\bibfield  {journal} {\bibinfo  {journal} {Euro. Phys. J.
  Spec. Top.}\ }\textbf {\bibinfo {volume} {226}} (\bibinfo {year}
  {2017})}\BibitemShut {NoStop}%
\bibitem [{\citenamefont {Levich}\ and\ \citenamefont
  {Krylov}(1969)}]{Levich1969}%
  \BibitemOpen
  \bibfield  {author} {\bibinfo {author} {\bibfnamefont {V.}~\bibnamefont
  {Levich}}\ and\ \bibinfo {author} {\bibfnamefont {V.}~\bibnamefont
  {Krylov}},\ }\href@noop {} {\bibfield  {journal} {\bibinfo  {journal} {Annu.
  Rev. Fluid Mech.}\ }\textbf {\bibinfo {volume} {1}},\ \bibinfo {pages} {293}
  (\bibinfo {year} {1969})}\BibitemShut {NoStop}%
\bibitem [{\citenamefont {Podlubny}(2001)}]{Podlubny2001}%
  \BibitemOpen
  \bibfield  {author} {\bibinfo {author} {\bibfnamefont {I.}~\bibnamefont
  {Podlubny}},\ }\href {http://arxiv.org/abs/math/0110241} {\bibfield
  {journal} {\bibinfo  {journal} {Fract. Calc. Appl. Anal.}\ }\textbf {\bibinfo
  {volume} {5}},\ \bibinfo {pages} {18} (\bibinfo {year} {2001})}\BibitemShut
  {NoStop}%
\bibitem [{\citenamefont {Atanackovi{\'{c}}}\ \emph {et~al.}(2014)\citenamefont
  {Atanackovi{\'{c}}}, \citenamefont {Pilipovi{\'{c}}}, \citenamefont
  {Stankovi{\'{c}}},\ and\ \citenamefont {Zorica}}]{Atanackovic2014}%
  \BibitemOpen
  \bibfield  {author} {\bibinfo {author} {\bibfnamefont {T.~M.}\ \bibnamefont
  {Atanackovi{\'{c}}}}, \bibinfo {author} {\bibfnamefont {S.}~\bibnamefont
  {Pilipovi{\'{c}}}}, \bibinfo {author} {\bibfnamefont {B.}~\bibnamefont
  {Stankovi{\'{c}}}}, \ and\ \bibinfo {author} {\bibfnamefont {D.}~\bibnamefont
  {Zorica}},\ }\href {\doibase 10.1002/9781118909065} {\emph {\bibinfo {title}
  {Fractional Calculus with Applications in Mechanics}}}\ (\bibinfo
  {publisher} {Wiley-ISTE},\ \bibinfo {year} {2014})\BibitemShut {NoStop}%
\bibitem [{\citenamefont {Miles}(1977)}]{Miles1977}%
  \BibitemOpen
  \bibfield  {author} {\bibinfo {author} {\bibfnamefont {J.}~\bibnamefont
  {Miles}},\ }\href {\doibase 10.1017/S0022112077001104} {\bibfield  {journal}
  {\bibinfo  {journal} {J. Fluid Mech.}\ }\textbf {\bibinfo {volume} {83}},\
  \bibinfo {pages} {153} (\bibinfo {year} {1977})}\BibitemShut {NoStop}%
\bibitem [{\citenamefont {Zhang}\ and\ \citenamefont
  {Vi{\~{n}}als}(1997)}]{Zhang1997}%
  \BibitemOpen
  \bibfield  {author} {\bibinfo {author} {\bibfnamefont {W.~B.}\ \bibnamefont
  {Zhang}}\ and\ \bibinfo {author} {\bibfnamefont {J.}~\bibnamefont
  {Vi{\~{n}}als}},\ }\href {\doibase 10.1017/S0022112096004764} {\bibfield
  {journal} {\bibinfo  {journal} {J. Fluid Mech.}\ }\textbf {\bibinfo {volume}
  {336}},\ \bibinfo {pages} {301} (\bibinfo {year} {1997})}\BibitemShut
  {NoStop}%
\bibitem [{\citenamefont {Landau}\ and\ \citenamefont
  {Lifshitz}(1966)}]{Landau1966}%
  \BibitemOpen
  \bibfield  {author} {\bibinfo {author} {\bibfnamefont {L.}~\bibnamefont
  {Landau}}\ and\ \bibinfo {author} {\bibfnamefont {E.}~\bibnamefont
  {Lifshitz}},\ }\href@noop {} {\emph {\bibinfo {title} {Fluid Mechanics}}},\
  \bibinfo {edition} {3rd}\ ed.\ (\bibinfo  {publisher} {Pergamon Press},\
  \bibinfo {year} {1966})\BibitemShut {NoStop}%
\bibitem [{\citenamefont {Edwards}\ \emph {et~al.}(1991)\citenamefont
  {Edwards}, \citenamefont {Brenner},\ and\ \citenamefont
  {Wasan}}]{Edwards1991}%
  \BibitemOpen
  \bibfield  {author} {\bibinfo {author} {\bibfnamefont {D.~A.}\ \bibnamefont
  {Edwards}}, \bibinfo {author} {\bibfnamefont {H.}~\bibnamefont {Brenner}}, \
  and\ \bibinfo {author} {\bibfnamefont {D.~T.}\ \bibnamefont {Wasan}},\
  }\href@noop {} {\emph {\bibinfo {title} {{Interfacial transport processes and
  rheology}}}}\ (\bibinfo  {publisher} {Butterworth-Heinemann},\ \bibinfo
  {year} {1991})\BibitemShut {NoStop}%
\end{thebibliography}%

\end{document}